\documentclass[journal]{IEEEtran}
\ifCLASSINFOpdf
\else
\fi
\usepackage{bbm}
\usepackage{verbatim}
\usepackage{graphicx}
\usepackage{cite}
\usepackage{url}
\usepackage[cmex10]{amsmath}
\usepackage{amssymb}
\usepackage{amsmath}
\usepackage{algorithm, algorithmicx, algpseudocode}
\usepackage[caption=false,font=footnotesize]{subfig}
\usepackage{color}
\usepackage{cite}
\usepackage{epstopdf}
\usepackage{calc}
\usepackage{array}
\usepackage{multirow}

\newtheorem{proposition}{Proposition}

\DeclareMathOperator*{\argmax}{arg\,max} 

\DeclareMathOperator{\diag}{\mathrm{diag}}

\newcommand{\timing}[0]{\text{E}}

\newcommand{\va}[1]{\mathbf{a}_{#1}}
\newcommand{\vA}[1]{\mathbf{A}_{#1}}

\newcommand{\tx}[0]{\text{T}}
\newcommand{\rx}[0]{\text{R}}

\newcommand{\hermitian}[0]{\text{H}}
\newcommand{\transpose}[0]{\text{T}}

\newcommand{\CFO}[0]{\epsilon_{\text{F}}}
\newcommand{\STO}[0]{\epsilon_{\text{T}}}
\newcommand{\Ts}[0]{T_{\text{s}}}
\newcommand{\Tb}[0]{T_{\text{B}}}
\newcommand{\Nb}[0]{N_{\text{B}}}
\newcommand{\Nc}[0]{N_{\text{c}}}
\newcommand{\Ncp}[0]{N_{\text{cp}}}
\newcommand{\sigman}[0]{\sigma_{\text{n}}}
\newcommand{\TSS}[0]{T_{\text{F}}}
\newcommand{\var}[0]{\mathrm{var}}
\newcommand{\Q}[0]{\mathrm{Q}}
\newcommand{\Qinv}[0]{\mathrm{Q}^{-1}}
\newcommand{\SNR}[0]{\mathrm{SNR}}
\newcommand{\Gr}[0]{G_{\text{R}}}
\newcommand{\Gt}[0]{G_{\text{T}}}
\newcommand{\Gd}[0]{G_{\text{D}}}
\newcommand{\prob}[0]{\mathrm{Pr}}
\newcommand{\NUE}[0]{N_{\text{U}}}

\makeatletter
\def\ps@IEEEtitlepagestyle{%
  \def\@oddfoot{\mycopyrightnotice}%
  \def\@evenfoot{}%
}
\def\mycopyrightnotice{%
  {\footnotesize This work has been submitted to the IEEE for possible publication. Copyright may be transferred without notice, after which this version may no longer be accessible.\hfill}
  \gdef\mycopyrightnotice{}
}

\begin{document}
%
\title{Compressive Initial Access and Beamforming Training for Millimeter-Wave Cellular Systems}

\author{Han~Yan,~\IEEEmembership{Student~Member,~IEEE},~and~Danijela~Cabric,~\IEEEmembership{Senior~Member,~IEEE}%
\thanks{Han Yan and Danijela Cabric are with the Electrical and Computer Engineering Department, University of California, Los Angeles, Los Angeles, CA 90095 (e-mail: yhaddint@ucla.edu; danijela@ee.ucla.edu).}
\thanks{Part of work was presented in IEEE GlobalSIP 2016 \cite{hyan_mmWave_CFO}}
\thanks{This work was supported by part under NSF grant 1718742.}
}



\maketitle


\begin{abstract}
Initial access (IA) is a fundamental physical layer procedure in cellular systems where user equipment (UE) detects nearby base station (BS) as well as acquire synchronization. Due to the necessity of using antenna array in millimeter-wave (mmW) IA, the channel spatial information can also be inferred. The state-of-the-art directional IA (DIA) uses sector sounding beams with limited angular resolution, and thus requires additional dedicated radio resources, access latency and overhead for refined beam training. To remedy the problem of access latency and overhead in DIA, this work proposes to use a quasi-omni pseudorandom sounding beam for IA, and develops a novel algorithm for joint initial access and fine resolution initial beam training without requiring extra radio resources. We provide the analysis of the proposed algorithm miss detection rate under synchronization error, and further derive  Cram\'er-Rao lower bound of angular estimation under frequency offset. Using QuaDRiGa simulator with mmMAGIC model at 28 GHz, the numerical results show that the proposed approach is advantageous to DIA with hierarchical beam training. The proposed algorithm offers up to two order of magnitude access latency saving compared to DIA, when the same discovery, post training SNR, and overhead performance are targeted. This conclusion holds true in various propagation environments and 3D locations of a mmW pico-cell with up to 140m radius. 
\end{abstract}


\section{Introduction}
\label{sec:Introduction}

\IEEEPARstart{T}HE millimeter-wave (mmW) communication is a promising technology for the future cellular network including 5G New Radio (5G-NR) \cite{Andrew:what5Gbe}. Due to abundant spectrum, it is expected that the mmW network will support ultra-fast data rate. As shown in both theory and prototypes, mmW system requires beamforming (BF) with large antenna arrays at both base station (BS) and user equipment (UE) to combat severe propagation loss \cite{Rappaport:mmWavewillwork}. 
Significant differences in propagation characteristics and hardware architectures for mmW band compared to microwave band require novel signal processing techniques \cite{7400949} and physical layer procedures \cite{3GPP_PHY_study}.

Initial access (IA) is the fundamental physical layer procedure that allows UE to discover and synchronize with nearby BS before further communication. However, IA for mmW networks brings new challenges and opportunities as compared to IA for sub-6GHz band networks. 
In mmW system, conventional omni-directional IA with single antenna can not be reliable, and as a result IA needs to leverage transmitter and receiver antenna array to exploit BF gain \cite{7161389,Giordani_beam_turotial_arxiv_1804}. A key design challenge in mmW IA is the design of sounding beams for reliable discovery.
In addition, beam training is required to achieve high BF gain enabled by large arrays and establish communication link. However, beam training now introduces additional access latency and signaling overhead due to repeated channel probing.
\subsection{Related works}

A number of works investigated various sounding beam designs and signal processing algorithms for mmW IA and beam training. Directional beams for IA and beam training are the most popular and extensively investigated in recent literature \cite{7161389,Caire_DIA_arxiv_1709,UTA_stocha_geo_IA_TCOM_18,KTH_stoch_geo_IA_arxiv_1802,Meng_OIA_TCOM_1803,Zorzi_IA_MCOM_1611,UTA_IA_TWC_1705,Giordani_beam_turotial_arxiv_1804,6847111,7914759}. Directional IA (DIA) is first studied in \cite{7161389} where a Generalized Likelihood Ratio Test (GLRT) is proposed to the solve the cell discovery problem under unknown multiple-input multiple-output (MIMO) channel and synchronization parameters. The authors concluded that the directional IA signal improves discovery range as compared to omni-directional IA. The DIA is further investigated in \cite{Caire_DIA_arxiv_1709} where overhead and access latency are analyzed. Works \cite{UTA_stocha_geo_IA_TCOM_18} and \cite{KTH_stoch_geo_IA_arxiv_1802} study DIA and its access latency in large networks using stochastic geometry. Impact of beam-width of sounding beams in DIA is researched in \cite{UTA_IA_TWC_1705}. The comparison between omni-directional and DIA is also discussed in \cite{Meng_OIA_TCOM_1803}. IA using out-of-band information, e.g., location, sub-6GHz measurement, are discussed in \cite{Zorzi_IA_MCOM_1611,Giordani_beam_turotial_arxiv_1804}. The aforementioned works mostly focused on the overhead and latency for the cell discovery, while beam training is either not discussed or assumed to have coarse resolution \cite{Caire_DIA_arxiv_1709}. It is common that DIA is paired with directional beam training \cite{6847111,7914759} where hierarchical sounding beams are used in multiple stages to achieve fine angular resolution for each user individually. However, such user-specific hierarchical sounding beams introduce prohibitive latency when a BS is connected to large number of UEs.

The alternative approaches for beam training are based on parametric  channel estimation \cite{7390019,7178503,8306126,7472310,8356247,8488662,8323164,7914742}. Exploiting the mmW sparse scattering nature, compressive sensing (CS) approaches have been considered to effectively estimate channel parameters based on channel observations obtained via various sounding beams. Works \cite{7390019,7178503} proposed a CS-based narrowband BF training with pseudorandom sounding beamformers in the downlink, and \cite{8306126} extended this approach for a wideband channel. Other related works include channel covariance estimation \cite{7472310,8356247,8488662} which requires periodic channel observations, and UE centric uplink training \cite{8323164,7914742}. It is worth nothing that all recent works focus on channel estimation alone while assuming perfect cell discovery and synchronization. The 5G-NR frame structure that supports IA is rarely considered, and further the feasibility of joint initial access and CS-based beam training has not been investigated.

There are also recent works that consider some practical aspects of IA. For example, frequency offset robust algorithms in narrowband mmW beam training are reported in \cite{hyan_mmWave_CFO,Myers_CFO_SPAWC_1707,8309152}. There are several hardware prototypes that consider a practical approach of using received signal strength (RSS) in CS-based beam training.  Channel estimation problem without phase measurement is a challenging problem, which was solved via novel signal processing algorithms based on RSS matching pursuit \cite{Rasekh_noncoherentCS_ACM_2017}, Hash table \cite{Hassanieh:2018:FMW:3230543.3230581}, and sparse phase retrieval \cite{UCSB_noncoherentCS_arxiv_1801}. Note that phase free measurements were associated with a particular testbed, and this constraint does not necessarily apply to mmW systems in general. In summary, while IA and beam training algorithms have been extensively studied in the literature, there is a lack of understanding about the theoretical limits and signal processing algorithms that jointly achieve cell discovery and accurate BF training using asynchronous IA signal in mmW frequency selective channel.

\subsection{Contributions}
In this work, we propose to use quasi-omni pseudorandom sounding beams and novel signal processing algorithm to jointly achieve initial cell discovery, synchronization, and fine resolution beam training. More specifically, we provide answers to the following questions.

\textit{How to use pseudorandom sounding beams for IA?} We propose an energy detection algorithm for initial discovery tailored for pseudorandom sounding beams. We derive the optimal detection threshold, analyze the miss detection probability and the impact of synchronization errors, i.e., carrier frequency offset (CFO) and timing offset (TO).

\textit{How to reuse received IA signal for beam training?} We propose a novel CS-based beam training algorithm that re-processes the frequency asynchronous IA signals to provide well aligned beam pair. We derive the Cram\'er-Rao lower bound (CRLB) of asynchronous training in line-of-sight (LOS) channel. We show that proposed algorithm reaches CRLB in LOS and remains effective in non-LOS (NLOS). 

\textit{What are the benefits of compressive IA?} We compare the proposed approach with DIA followed by hierarchical directional beam training. Key performance indicators for both approaches are numerically compared, including discovery rate, post beam training SNR, overhead and access latency. The simulation study based on 5G-NR frame structure and measurement-endorsed 3D 28GHz channel shows that the proposed approach is advantageous to DIA for UEs across wide range of locations in a small cell.



\subsection{Organizations and notations}
The rest of the paper is organized as follows. We start with a brief introduction of 5G-NR frame structure, IA and beam training in Section~\ref{sec:preliminary}. In Section~\ref{sec:system_model}, we present the system model and problem statement. Section~\ref{sec:cell_discovery} includes the proposed algorithm for cell discovery and timing acquisition followed by associated performance analysis. In Section~\ref{sec:CS_based_training} we present the algorithm and analysis for initial beam training under CFO. The access latency, overhead, and complexity analysis is included in Section~\ref{sec:comparison_analysis}. The numerical results are presented in Section~\ref{sec:simulation results}. Open research issues are summarized in Section~\ref{sec:discussion}. Finally, Section~\ref{sec:Conclusion} concludes the paper.


\textit{Notations:} Scalars, vectors, and matrices are denoted by non-bold, bold lower-case, and bold upper-case letters, respectively.
The $(i,j)$-th element of $\mathbf{A}$ is denoted by $[\mathbf{A}]_{i,j}$. Conjugate, transpose, Hermitian transpose, and pseudoinverse are denoted by $(.)^{*}$, $(.)^{\transpose}$, $(.)^{\hermitian}$, and $(.)^{\dagger}$ respectively. The inner product is $\langle \mathbf{a},\mathbf{b}\rangle  \triangleq  \mathbf{a}^{\hermitian}\mathbf{b}$. The $l_2$-norm of $\mathbf{h}$ is denoted by $||\mathbf{h}||$. 
$\diag(\mathbf{a})$ aligns vector $\mathbf{a}$ into a diagonal matrix. Kronecker and Hadamard product are denoted as $\otimes$ and $\circ$, respectively. $\Re(x)$ and $\Im(x)$ are the real and imaginary parts of $x$, respectively. Set $\mathcal{S}=[a,b]$ contains all integers between $a$ and $b$.


\section{Preliminaries: initial access and beam training}
\label{sec:preliminary}
In this section, we introduce the mmW physical layer initial access procedure in 5G-NR cellular network. We briefly review the frame structure, synchronization sequences, and directional IA scheme as well as beam training. The reader is referred to work \cite{Giordani_beam_turotial_arxiv_1804} for a more detailed survey.
\begin{figure}
\begin{center}
\includegraphics[width=0.48\textwidth]{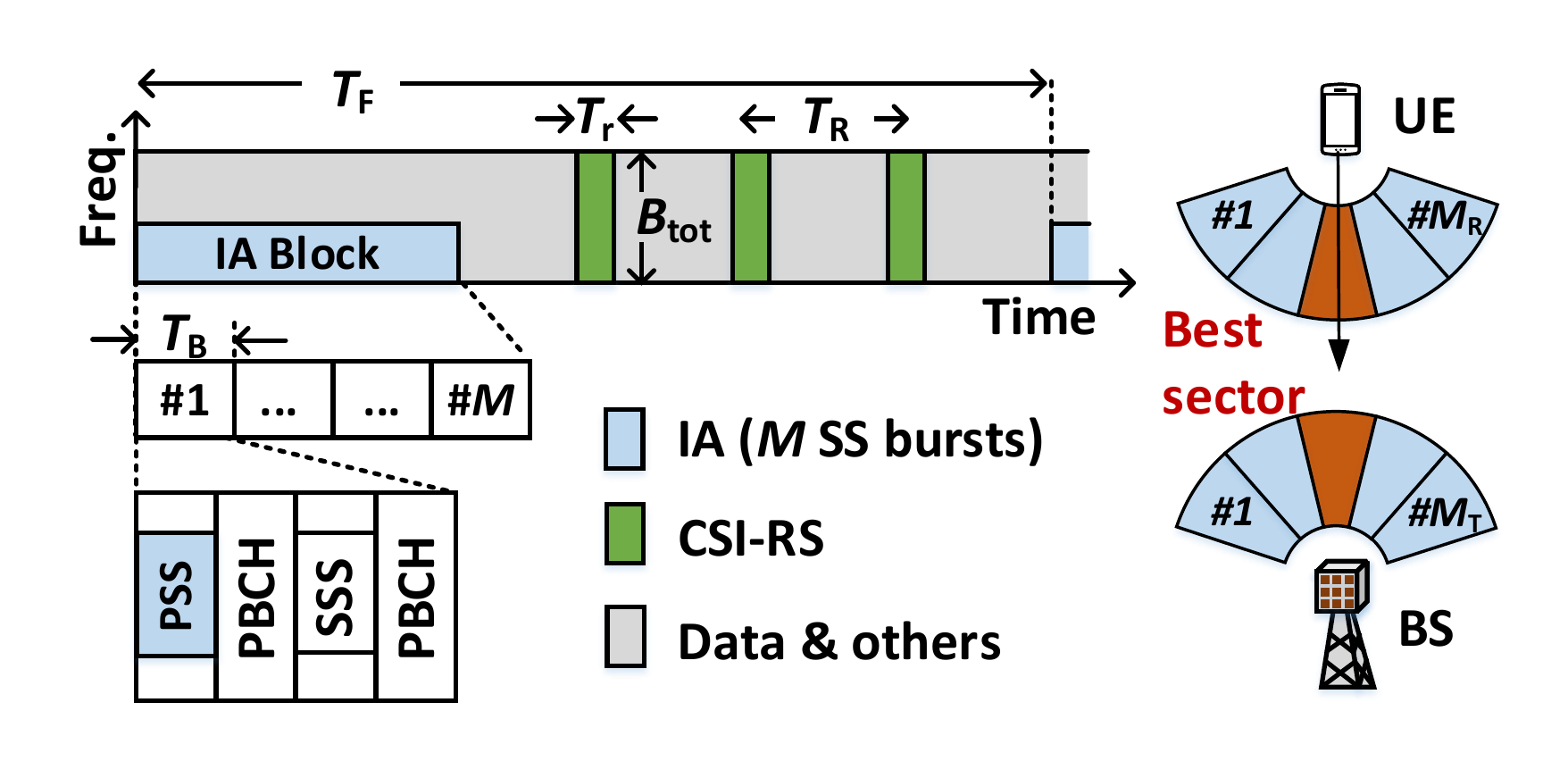}
\end{center}
\vspace{-4mm}
\caption{The 5G-NR mmW frame structure with emphasis in beam management function and the illustration for directional initial access.}
\vspace{-4mm}
\label{fig:frame_structure}
\end{figure}

\textit{Frame Structure:} Fig.~\ref{fig:frame_structure} shows the frame structure of 5G-NR. We focus on two functional blocks, namely synchronization signal (SS) burst and channel state information reference signal (CSI-RS). 5G-NR uses orthogonal frequency division multiplexing (OFDM), and the subcarrier spacing is either $120 \text{ or } 240$ KHz for mmW band. The SS signal is transmitted by a BS with period $\TSS$, typically 20 ms. The SS consists of up to $M=64$ burst blocks. In each one of the burst blocks of duration $T_{\text{B}}$, a specific sounding beam pair is used by BS and UE. The CSI-RS block with duration $T_{\text{r}}$ is dedicated to specific UE(s) for beam training and tracking. CSI-RS can use all frequency resources, i.e., up to  $B_{\text{tot}}$, and it has periodicity of $T_{\text{R}}$, an implementation dependent value.

\textit{Synchronization Signal:}
Referring to Fig.~\ref{fig:frame_structure}, each SS burst has 4 OFDM symbols, i.e., primary synchronization signal (PSS), physical broadcast channel (PBCH), and secondary synchronization signal (SSS), followed by another PBCH. PSS is used in cell detection and synchronization, and it is assigned to the middle $P=128$ subcarriers of the first OFDM symbol. \color{black} The PSS in 4G-LTE is based on Zadoff-Chu (ZC) sequences due to their perfect cyclic-autocorrelation property and their Fourier duals \cite{4373412}, while in 5G-NR PSS is replaced by Maximum Length Sequences (M-sequences)  \cite{3GPP_modulation}. There are $N_{\text{PPS}} = 3$ and 336 unique sequences of PSS and SSS, respectively, and these 1008 combination define the cell identifier (ID) of BS. PBCH carries control information.
\color{black}

\textit{Beamformed Initial Access:}
The BS periodically transmits IA blocks and such signals are processed by UEs which desire to establish the initial access, reconnect after beam misalignment, and search for additional BSs for potential handover. The sounding beams in SS bursts are intended to facilitate multi-antenna processing in BS and UE when no a priori channel information is available. Referring to Fig.~\ref{fig:frame_structure}, BS and UE in the DIA scheme use $M_{\tx}$ and $M_{\rx}$ transmitter and receiver beams to cover angular space at both ends. One T/Rx beam is used at a time, for all $M=M_{\tx}M_{\rx}$ SS bursts. 



\textit{Beam Training:} 
The purpose of beam training is to identify the best beam pairs between BS and UE. The sounding beams in DIA typically have large beam-width and flat response inside angular sectors \cite{7845674}. Such design covers the angular space of BS and UE within $M$ bursts, but achieves coarse propagation directions estimation \cite{Caire_DIA_arxiv_1709}. Thus DIA relies on directional beam training to refine angular resolution where BS and UE steer narrow sounding beams within the sectors of interest during CSI-RS periods.


%
%
\section{System Model}
\label{sec:system_model}

This section introduces the system model that adopts the 5G-NR frame structure and problem formulation. All important notations are summarized in Table~\ref{tab:notation}.
\begin{table}
\caption{Nomenclature}
\centering
\begin{tabular}{|c|c|}
\hline 
Symbol &Explanations\tabularnewline
\hline 
\hline 
$p$, $P$ &  Index and total number of subcarriers\tabularnewline
$m$, $M$ &  Index and total number of SS bursts\tabularnewline
$l, L$  & Index and total number of multipaths \tabularnewline
\hline 
$N_{\tx}$, $N_{\rx}$  & Number of antenna in BS and UE\tabularnewline
$\Ts$ &  Sample duration of IA signal\tabularnewline 
$\Tb,\Nb$ &  Duration and sample number in each SS burst\tabularnewline
$\TSS$ &  Period of SS bursts\tabularnewline
$T_{\text{R}}$,$T_{\text{r}}$ &  Period and duration of CSI-RS\tabularnewline
$\Nc$, $\Ncp$ &  Max. excess delay taps and length of CP\tabularnewline
$N_{\text{train}},N_{\text{U}}$ &  Required CSI-RS and UE number\tabularnewline
\hline 
$\Delta f$, $\epsilon_{\text{F}}$ &  CFO in [Hz] and normalized in [rad/samp.] \tabularnewline
$\epsilon_{\text{T}}$ &  Initial TO in UE (number of sample) \tabularnewline
\hline 
$\mathbf{H}[d]$ & MIMO channel at $d$-th delay sample\tabularnewline
$\mathbf{a}_{\tx}\left(\theta\right)$, $\mathbf{a}_{\rx}\left(\phi\right)$   & Spatial responses of BS and UE\tabularnewline
$\phi_l,\theta_l$, $g_l$, $\tau_l$  &Gain/AoA/AoD/delay of $l$-th multipath  \tabularnewline
$\alpha_l, \beta_l$  & Real and imaginary parts of $g_l$  \tabularnewline
\hline 
$\mathbf{s}, \tilde{\mathbf{s}}$, $\tilde{s}[n]$ & F/T domain PSS vector and sequence\tabularnewline
$\mathbf{v}_m, \mathbf{w}_m$ & RF precoder/combiner of the $m$-th burst\tabularnewline
$z[n]$, $\mathbf{z}_m$, $\sigman^2$ & AWGN sequence, vector, and power\tabularnewline
\hline 
\multicolumn{2}{|c|}{Initial discovery (detection)}\tabularnewline
\hline
$P^{\star}_{\text{FA}}$ & Target FA prob. in initial discovery \tabularnewline
$P_{\text{MD,PT}}$, $P_{\text{MD,NT}}$ & MD prob. w/ and w/o perfect timing  \tabularnewline
$\gamma_{\text{PT}}, \eta_{\text{PT}}$ & Detection stat. and TH w/ perfect timing\tabularnewline
$\gamma_{\text{NT}}, \eta_{\text{NT}}$ & Detection stat. and TH w/ unknown timing\tabularnewline
\hline 
\multicolumn{2}{|c|}{Initial beamforming traning (estimation)}\tabularnewline
\hline
$\boldsymbol{\xi}$ & Unknown parameters in BF training \tabularnewline
$\mathbf{y}_m$  & Received OFDM symbols at $m$-th burst \tabularnewline
$\mathbf{d}$, $\mathbf{t}$, $\mathbf{r}$ & Vectors with candidates delay/AoA/AoD \tabularnewline
$\Gd$, $\Gt$, $\Gr$ & Parameter grid in delay/AoA/AoD est. \tabularnewline
$\mathbf{Q}$, $\mathbf{F}$ & ICI matrix and DFT matrix \tabularnewline

\hline
\end{tabular}
\label{tab:notation}
\end{table}

%
%
\subsection{Received signal model before timing acquisition}
\label{sec:signal_model_before_timing}
Consider a single cell system with a BS equipped with $N_{\tx}$ antennas. The BS transmits beamformed IA signal over mmW sparse multipath channel to UEs. We focus on the IA and BF training procedure for a single UE. It is straightforward to extend it to multiple UEs since there is no UE-dependent processing. The UE uses analog array architecture, i.e., phased array, with $N_{\rx}$ antennas. We assume that a single stream of IA signal is transmitted by the BS regardless of its architecture.
\begin{figure}
\begin{center}
\includegraphics[width=0.48\textwidth]{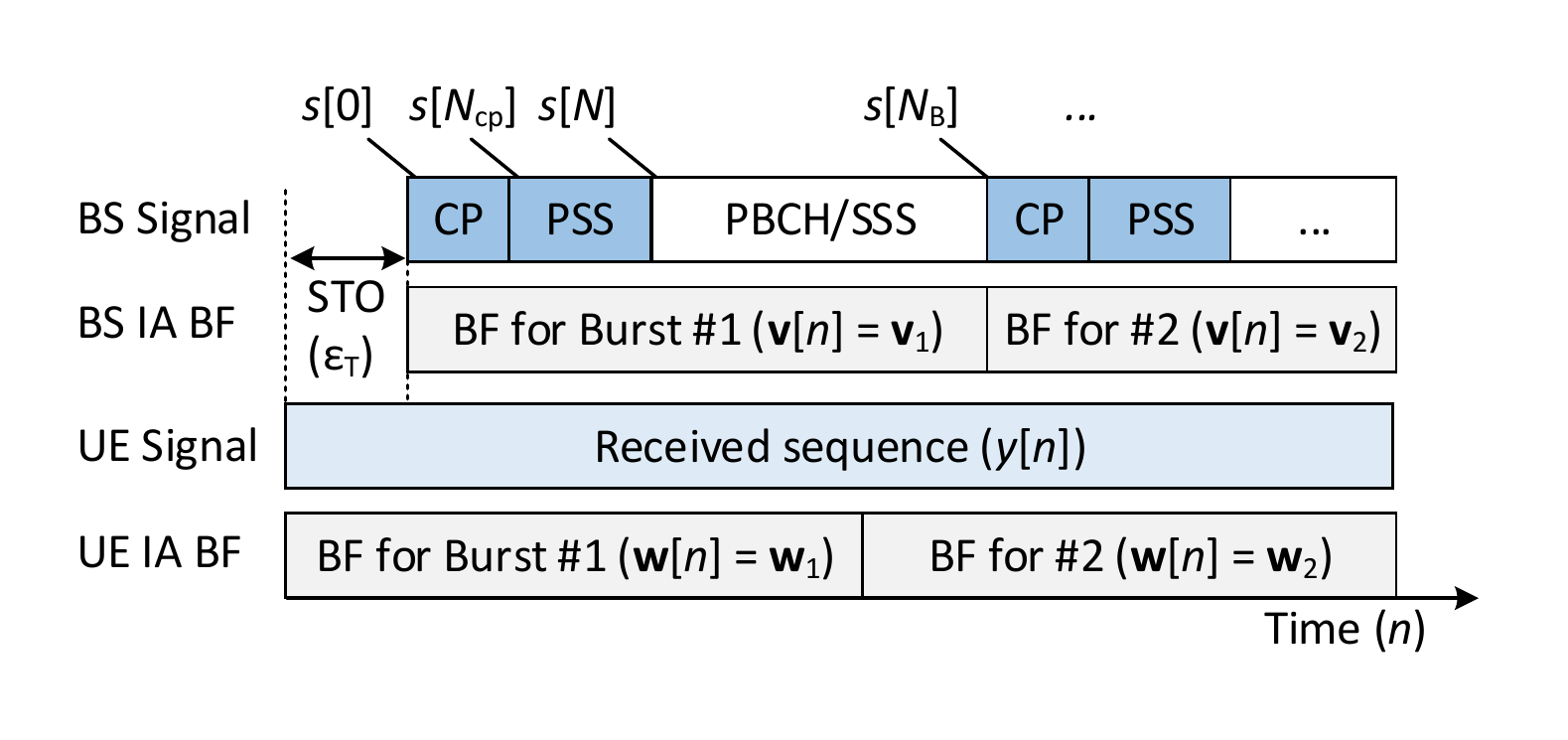}
\end{center}
\vspace{-5mm}
\caption{Illustration of the received signal model as time sequence.}
\vspace{-5mm}
\label{fig:system_mode}
\end{figure}

We first consider the received signal model when a UE searches for BS to initialize the connection. In this procedure, UE follows a periodic SS burst structure and uses predefined receiver beamformers to capture the signal according to \cite{Giordani_beam_turotial_arxiv_1804}. As illustrated in Fig.~\ref{fig:system_mode}, when the signal is present, the received samples, sampled at $\Ts$, is denoted as
\begin{align}
\begin{split}
y[n] =  &\sum_{d=0}^{\Nc-1}e^{j(\CFO n+\psi[n])}\mathbf{w}^{\hermitian}[n]\mathbf{H}[d]\mathbf{v}[n-d-\epsilon_{\text{T}}]s[n-d-\epsilon_{\text{T}}]\\
&+ \mathbf{w}^{\hermitian}[n]\mathbf{z}[n],\quad  n\in [0,N_{\text{F}}-1].
\end{split}
\label{eq:rx_signal_raw}
\end{align}
In the above equation, $\STO$ is the unknown integer sample TO within range\footnote{We assume coarse timing synchronization is available with 10 $\mu$s level accuracy that corresponds to current LTE-A. Practically it is achievable via GPS clock or non-standalone mmW network \cite{Giordani_beam_turotial_arxiv_1804}.} $0\leq \STO \leq \epsilon_{\text{T,max}} \leq \Nb$, where $\epsilon_{\text{T,max}}$ is the largest offset known to the system and $\Nb$ is the number of samples in one SS burst, i.e., $N_{\text{B}}=\Tb/\Ts$. The phase measurement error $e^{j(\CFO n+\psi[n])}$ comes from two sources. $\CFO$ is the normalized initial CFO, i.e., $\CFO = 2\pi\Ts \Delta f$ where $\Delta f$ is absolute CFO in Hz between BS and UE. $\psi[n]$ is the phase noise process in the UE receiver.
$\Nc$ is the maximum excessive multipath delay in discrete time, based on which cyclic prefix (CP) $\Ncp>\Nc$ for OFDM symbols is designed. $s[n]$ is the time domain signals of SS bursts.
%
Referring to Fig.~\ref{fig:system_mode}, we focus on the PSS and treat other symbols as zero \cite{7161389}, i.e., 
\begin{align*}
s[n]=
\begin{cases}
s_{\text{zc}}[n-(m-1)N_{\text{B}} + P - N_{\text{CP}}], & n \in \mathcal{S}_{\text{CP},m}\\
s_{\text{zc}}[n-(m-1)N_{\text{B}}-N_{\text{CP}}], & n \in \mathcal{S}_{\text{PSS},m}\\
0, & \text{otherwise}
\end{cases},
\end{align*}
where
$\mathcal{S}_{\text{CP},m}\triangleq [(m-1)N_{\text{B}},(m-1)N_{\text{B}}+N_{\text{CP}}-1]$, $\mathcal{S}_{\text{PSS},m}\triangleq [(m-1)N_{\text{B}}+N_{\text{CP}},(m-1)N_{\text{B}}+N-1]$
are the sets with sample index corresponding to CP and PSS in the $m$-th burst, respectively. $|s_{\text{zc}}[n]|=1, n\in [0,P-1]$ is the Fourier dual of a known PSS sequence, and $N=P+\Ncp$ is the number of samples in PSS including CP.
%
%
$\mathbf{z}[n]$ is the Additive White Gaussian noise (AWGN) and $\mathbf{z}[n]\sim \mathcal{CN}(0,\sigman^2\mathbf{I}_{N_{\rx}})$. Vectors $\mathbf{v}[n]$ and $\mathbf{w}[n]$ are beamformers used by BS and UE at instance $n$, respectively, and they are from a predefined set of IA beam codebook, i.e., $\mathbf{w}[n] \in \mathcal{W}\triangleq \{\mathbf{w}_1,\cdots,\mathbf{w}_M\}$ and $\mathbf{v}[n] \in \mathcal{V}\triangleq \{\mathbf{v}_1,\cdots,\mathbf{v}_M\}$. BS and UE sequentially use respective beamformers for an interval of $\Nb$ samples and switch to the next one in $\mathcal{W}$ and $\mathcal{V}$, i.e.,
$\mathbf{w}[n] =  \mathbf{w}_m, \text{ if }\lfloor n/\Nb\rfloor = m$ and $\mathbf{v}[n] =  \mathbf{v}_m, \text{ if }\lfloor n/\Nb\rfloor = m.$ Beamformer switching is assumed not to introduce latency or phase offset in the transmission and reception.
In this work, we focus on the system where each element of $\mathbf{v}_m$ and $\mathbf{w}_m$ is randomly and independently chosen from a set
$\mathcal{S}_{\tx} = \left\{\pm 1/\sqrt{N_{\tx}},\pm j\sqrt{N_{\tx}}\right\},$ and
$\mathcal{S}_{\rx} = \left\{\pm 1\sqrt{N_{\rx}},\pm j\sqrt{N_{\rx}}\right\}.$
Such sounding beams require only 4-level phase quantization when steered by phased array and have randomized quasi-omnidirectional beam pattern.

The discrete time MIMO channel at delay $d$ ($d<N_{\text{c}}$) is denoted as $\mathbf{H}[d]\in \mathbb{C}^{N_{\rx}\times N_{\tx}}$. 
Following the extended Saleh Valenzuela (S-V) model in \cite{7400949}, we express $\mathbf{H}[d]$ as 
\begin{align*}
\begin{split}
\mathbf{H}[d] = &\frac{1}{\sqrt{N_{\tx}N_{\rx}}}\sum_{l=1}^{L}\sum_{r=1}^{R} g_{l,r} p_{\text{c}}(d\Ts-\tau_{l,r})\va{\rx}(\phi_{l,r}) \va{\tx}^{\hermitian}(\theta_{l,r}),
\end{split}
\end{align*}
where $L$ and $R$ are the number of multipath clusters (typically small, $L\leq 4$ \cite{7109864}) and sub-paths (rays), respectively. Scalar $g_{l,r}$, $\tau_{l,r}$, $\theta_{l,r}$ and $\phi_{l,r}$ are the complex gain, excessive delay, angle of departure (AoD) and angle of arrival (AoA) of the $r$-th sub-path within the $l$-th cluster, respectively. \color{black}Function $p_{\text{c}}(t)$ is the time domain response filter due to limited temporal resolution $\Ts$. \color{black} 
With half wavelength antenna spacing, the angular response vectors at the BS and UE are denoted as $\va{\tx}(\theta)\in\mathbb{C}^{N_{\tx}}$ and $\va{\rx}(\phi)\in\mathbb{C}^{N_{\rx}}$. Their defined $k$-th element is 
$[\mathbf{a}_{\rx}(\phi)]_{k}  = \text{exp}[j\pi(k-1)\sin(\phi)]$ and $[\mathbf{a}_{\tx}(\theta)]_{k}  = \text{exp}[j\pi(k-1)\sin(\theta)]$.

\color{black}Note that the above model aligns with measurement-endorsed mmMAGIC channel model \cite{mmMAGIC_model} and is used for the system performance evaluation in Section~\ref{sec:simulation results}. However, for the sake of tractable algorithm design and analysis, the following assumptions and definitions are made.\color{black}

\textit{Assumption 1:} BS and UE have ULA with omni-directional element pattern in 2D environment. Intra-cluster AoA, AoD, and delay offsets are zero, i.e., $\sum_{r=1}^{R}g_{l,r} \triangleq g_l, \phi_{l,r} = \phi_{l}$, $\theta_{l,r}=\theta_{l}$, $\tau_{l,r} =\tau_{l}, \forall r$. Index $r$ is omitted in the rest of paper for clarity. The phase error process is solely from CFO i.e., phase noise process is $\psi[n]=0,\forall n$ in (\ref{eq:rx_signal_raw}). The complex path gain $g_l$ is deterministic complex value, i.e., $\sum_{l=1}^{L}|g_l|^2=\sigma^2_{\text{g}}$. 

\textit{Definition 1:}
The pre-BF signal to noise ratio (SNR) is defined as 
$\SNR \triangleq\sigma^2_{\text{g}}/\sigman^2$.



%
%
\subsection{Problem formulations}

We intend to address the following two problems, and their connection to the existing works are remarked.

\textit{Problem 1 (Initial Discovery and Timing Acquisition):} The UE needs to detect the SS burst from in-band received samples  (\ref{eq:rx_signal_raw}). This problem is a binary hypothesis testing with unknown channel $\mathbf{H}[d]$ and synchronization errors $\STO$ and $\CFO$.

\begin{align}
\begin{split}
\mathcal{H}_0: \quad & y[n] = \mathbf{w}^{\hermitian}[n]\mathbf{z}[n],\\
\mathcal{H}_1:\quad & y[n] =  \sum_{d=0}^{N_\text{c}-1}\bigg(e^{j\CFO n}\mathbf{w}^{\hermitian}[n]\mathbf{H}[d]\mathbf{v}[n-d-\epsilon_{\text{T}}]\\
&\quad\quad\quad \cdot s[n-d-\epsilon_{\text{T}}]\bigg)+ \mathbf{w}^{\hermitian}[n]\mathbf{z}[n].
\end{split}
\end{align}
In addition, the TO $\STO$ is estimated at this stage.

\textit{Problem 2 (Initial BF Training):} The BF training is triggered once UE has detected IA signals. In this stage, UE re-uses the asynchronous signal samples (\ref{eq:rx_signal_raw}) to estimate the AoD and AoA of a path with significant power, say $\theta^{\star}$ and $\phi^{\star}$, and which are then used as the steering vectors, $\mathbf{v}^{\star} = \mathbf{a}_{\tx}(\theta^{\star})$ and $\mathbf{w}^{\star} = \mathbf{a}_{\rx}(\phi^{\star})$, in data communications phase. 

\textit{Remark 1:} The above problems can be solved by DIA and directional beam training with the help of CSI-RS, while our solution relies on processing IA block only. In additional, although \textit{Problem 2} has overlap with parametric channel estimation, approaches from this class are not directly comparable. In fact, \cite{8306126,8323164} estimate the entire wideband channel, which facilitates optimal MIMO processing, but the assumptions of perfect synchronization and capability of wideband probing, i.e., $B_{\text{tot}}$, do not apply in our work. Our goal is to provide well-aligned beam pair within IA block, i.e. without requiring CSI-RS slots. Finally, the cell ID recognition and PBCH decoding are important tasks but are not studied in this work.

%
%
\section{Initial Discovery and Timing synchronization}
\label{sec:cell_discovery}
This section presents the proposed initial discovery and timing synchronization algorithm followed by their performance analysis.
\subsection{Initial discovery and timing synchronization algorithm}
The UE processes the received signal using the correlation filter with $s_{\text{zc}}[n]$, and obtains the detection statistics:
\begin{align}
\tilde{y}[n] = \frac{1}{P} \sum_{k=0}^{P-1}y[n+k]s_{\text{zc}}^*[k].
\label{eq:ZC_correlation}
\end{align}
Intuitively, there are $M$ correlation peaks across $M$ SS bursts. The magnitude of the $m$-th peaks depends on the array gain of the $m$-th sounding beamformer, CFO, and TO. Our proposed detector combines energy from all $M$ SS bursts and compares it with the threshold. In contrast to previous works \cite{7161389,KTH_stoch_geo_IA_arxiv_1802,UTA_stocha_geo_IA_TCOM_18} where the detection threshold is a fixed constant, we propose to use the optimal detection threshold based on Neyman-Pearson criterion that meets target false alarm (FA) rate $P^{\star}_{\text{FA}}$.

To understand the impact of timing synchronization error, we first consider a \textit{Genie} scenario where the UE has perfect timing (PT) information, i.e., $\STO = 0$. In this case, the proposed PSS detection scheme is an energy detector over all $M$ bursts. In addition, a sample time window with $\Nc$ is used to collect energy from all multipaths. Specifically, the proposed hypothesis testing scheme is expressed as
\begin{align}
\gamma_{\text{PT}} \triangleq \frac{1}{M}\sum_{m=0}^{M-1}\sum_{k=0}^{\Nc-1}\left|\tilde{y}[k+mN_{\text{b}}]\right|^2 \underset{\mathcal{H}_0}{\overset{\mathcal{H}_1}{\gtrless}} \eta_{\text{PT}},
\label{eq:ED_wo_STO}
\end{align}
where the detection threshold $\eta_{\text{PT}}$ is used to reach false alarm rate constraint such that $\prob(\gamma_{\text{PT}} > \eta_{\text{PT}}|\mathcal{H}_0) = P^{\star}_{\text{FA}}$.

In a practical scenario without initial timing information (NT), i.e., $\STO\neq 0$, we propose to use the following detector
\begin{align}
\gamma_{\text{NT}} \triangleq \max_{0\leq n < \epsilon_{\text{T,max}}} \frac{1}{M}\sum_{m=0}^{M-1}\sum_{k=0}^{\Nc -1}\left|\tilde{y}[n+k+mN_{\text{b}}]\right|^2 \underset{\mathcal{H}_0}{\overset{\mathcal{H}_1}{\gtrless}}  \eta_{\text{NT}}
\label{eq:ED_w_STO}
\end{align}
that searches all possible instances within TO window $\STO \in [0,\epsilon_{\text{T,max}}]$ and uses the highest energy collected for the hypothesis test.
%
The sample index corresponding to the highest energy in (\ref{eq:ED_w_STO}) is the estimate of TO, namely
\begin{align}
\hat{\epsilon}_{\text{T}} = \argmax_{0\leq n<\epsilon_{\text{T,max}}} \frac{1}{M}\sum_{m=0}^{M-1}\sum_{k=0}^{N_{\text{c}-1}}\left|\tilde{y}[n+k+mN_{\text{b}}]\right|^2.
\label{eq:timing_acquisition}
\end{align}



%
%
\subsection{Performance of initial discovery and timing acquisition}
In this subsection, we analyze performance of the proposed discovery algorithm in terms of miss detection rate, and  
the impact of initial synchronization error $\CFO$ and $\STO$. The exact expression is challenging and tedious, if not impossible, and therefore we provide a tight closed-form approximation in the following proposition. To be concise, the subscripts of $\gamma$ and $\eta$ that indicate the timing information assumption are denoted as binary variable $\timing \in \{\text{NT},\text{PT}\}$.

\begin{proposition}
\label{proposition:MD_w_STO}
The optimal threshold of (\ref{eq:ED_w_STO}) that reaches target FA rate $\prob(\gamma_{\text{\timing}} \geq \eta^{\star}_{\timing}|\mathcal{H}_0) = P^{\star}_{\text{FA}}$ is approximately\footnote{Approximation is tight when TO search window size $\epsilon_{\text{T,max}}\geq 100$.}
\begin{align}
\begin{split}
&\eta^{\star}_{\timing} = \sigman^2\left[\frac{\Nc}{P}+\sqrt{\frac{\Nc}{MP^2}}\xi_{\text{z}}\left(\epsilon_{\text{T,max}},P^{\star}_{\text{FA}}\right)\right],
\end{split}
\label{eq:TH_w_STO}
\end{align}
where $\xi_{\timing}(\epsilon_{\text{T,max}},P^{\star}_{\text{FA}})$ is the threshold adjustment factor dependent on synchronization computed as
\begin{align}
&\xi_{\timing} = \begin{cases}
\Qinv\left(P^{\star}_{\text{FA}}\right),& \timing=\text{PT}\\
\Qinv\left(\frac{1}{\epsilon_{\text{T,max}}}\right) - \frac{0.78\ln\left(-\ln\left(1-P^{\star}_{\text{FA}}\right)\right)}{\Qinv\left(\frac{1}{\epsilon_{\text{T,max}}}\right)},& \timing=\text{NT}
\end{cases},
\label{eq:xi_threshold_adjust}
\end{align} 
where $\Q(.)$ and $\Qinv(.)$ are Q-function and inverse Q-function, respectively. The associated miss detection (MD) rate $P_{\text{MD},\timing} \triangleq \prob\left(\gamma_{\timing} < \eta^{\star}_{\timing}|\mathcal{H}_1\right)$ using the optimal threshold $\eta^{\star}_{\timing}$ is
\begin{align}
\begin{split}
&P_{\text{MD},\timing} =
 \Q \left(\frac{\kappa(\STO,\CFO) \SNR -\sqrt{\frac{\Nc}{MP^2}}\xi_{\text{z}}\left(\epsilon_{\text{T,max}},P^{\star}_{\text{FA}}\right)}{\sqrt{\frac{2\kappa^2(\STO,\CFO)\SNR^2}{M}+ \frac{\Nc}{P^2M}}}\right),
\end{split}
\label{eq:PMD_w_STO}
\end{align}
where the SNR degradation factor $\kappa(\STO,\CFO)$ is defined as
\begin{align}
\kappa(\STO,\CFO) = \frac{2-\Re\left(e^{jK(\STO)\CFO}\right)-\Re\left(e^{j[P-K(\STO)]\CFO}\right)}{P^2\left[1-\Re\left(e^{j\CFO}\right)\right]},
\label{eq:kappa_SNR_degradation}
\end{align}
where $K(\STO)$ is the number of samples during PSS reception that UE switches beamformer due to TO.
\begin{align}
K(\STO)=
\begin{cases}
\Nb-\STO, &\text{if } \Nb-P \leq \STO < \Nb\\
0,     &\text{otherwise}
\end{cases}.
\label{eq:define_gain_split_K}
\end{align}

\end{proposition}
\begin{IEEEproof}
See Appendix \ref{appendix:detection_derivation_wo_STO}.
\end{IEEEproof}

\textit{Remark 2:} $1-P_{\text{MD},\text{NT}}$ is a close approximation of probability that UE detects IA and correctly estimates $\STO$.

We gain two main insights from MD expressions (\ref{eq:PMD_w_STO}) corresponding to threshold adjustment factor $\xi_{\text{z}}(\epsilon_{\text{T,max}},P^{\star}_{\text{FA}})$ and SNR degradation factor $\kappa(\CFO,\STO)$. Firstly, the CFO affects MD performance by effectively reducing SNR via term $\kappa(\CFO,\STO)$. Under maximum CFO at UE of $\pm 5$ppm and typical frame parameters $P,M,\Nc$ specified in Section~\ref{sec:simulation results}, the SNR degradation is bounded by 4 dB, i.e., $10\log_{10}[\kappa(\CFO,\STO)]\geq-4\text{dB},\forall \STO$. 
Secondly, the TO has impact on both factors. As seen in (\ref{eq:kappa_SNR_degradation}), the SNR in the detection problem degrades when severe TO exists. In fact, $K(\STO)$ in $\kappa(\CFO,\STO)$ models phenomenon that receiver sounding beam switches during the reception of PSS, i.e., $K(\STO)\neq 0$. In addition, the presence of TO forces system to use peak detection scheme (\ref{eq:ED_w_STO}) where system searches peak location over a sample window with length $\epsilon_{\text{T,max}}$, i.e.,the worst case in (\ref{eq:ED_w_STO}). Under $\mathcal{H}_0$, the algorithm picks strongest noise realization over $\epsilon_{\text{T,max}}$ samples and thus system needs to use higher threshold than in PT scenario, as seen in (\ref{eq:TH_w_STO}) and (\ref{eq:xi_threshold_adjust}). Note that such degradation does not depend on the value of $\STO$, and the degradation in (\ref{eq:PMD_w_STO}) is not critical with practical maximum TO uncertainty $\epsilon_{\text{T,max}}\leq \Nb$. In summary, synchronization offset does not severely affect discovery performance of the proposed scheme.
%
%



%
%
\subsection{Benchmark approach: directional initial discovery}
For completeness, we briefly introduce the benchmark approach using directional sounding beam in initial discovery \cite{7161389}. The system model of DIA is similar to Section~\ref{sec:system_model}, except that sounding beamformers $\mathcal{W}$ and $\mathcal{V}$ are codebooks that steer directional sector beams, e.g., \cite{6847111,array_textbook}. Adapting the approach in \cite{7161389} for the wideband channel and known PSS in SS burst, the cell discovery in DIA uses the following detector
\begin{align}
\gamma_{\text{DIA}} \triangleq \max_n \left|\tilde{y}_{\text{DIA}}[n]\right|^2 \underset{\mathcal{H}_0}{\overset{\mathcal{H}_1}{\gtrless}} \eta_{\text{DIA}}
\label{eq:DIA}
\end{align}
where $\gamma_{\text{DIA}}$ and $\gamma_{\text{DIA}}$ are the detection statistic and threshold in DIA. Sequence $\tilde{y}_{\text{DIA}}[n]$ is the correlation output in (\ref{eq:ZC_correlation}) that corresponds to directional sounding beams. Refer to Fig.~\ref{fig:frame_structure}, the UE detects the burst with maximum power and denotes the index as $m_{\text{DIA}}^{\star}$ which is used in directional beam training.
%
%
\section{Compressive Initial beam training}
\label{sec:CS_based_training}
This section presents the proposed initial access based BF training. We start with signal rearrangement based on information obtained from successful cell discovery and timing acquisition. Then, we introduce the CS problem formulation followed by the proposed algorithm. Finally, we analyze the CRLB of AoA/AoD estimation in LOS.

\subsection{Signal rearrangement after timing acquisition}
\label{sec:signal_rearrangement}

The further processing requires correct detection and CP removal, and therefore we make a following assumption.

\textit{Assumption 2:} In beam training, the received IA signal (\ref{eq:rx_signal_raw}) is correctly detected and TO $\STO$ is correctly estimated. 

The UE first removes CPs of $P$ PSS samples from $y[n]$ corresponding to $M$ bursts and rearranges them into vector 
\begin{align}
\begin{split}
\mathbf{y} = &[\mathbf{y}^{\transpose}_1,\cdots,\mathbf{y}^{\transpose}_m,\cdots,\mathbf{y}^{\transpose}_M]^{\transpose},\\
\{\mathbf{y}_m\}_p = &y[\hat{\epsilon}_{\text{T}}+N_{\text{CP}}+(p-1)+(m-1)\Nb], p\leq P.
\end{split}
\label{eq:rearrangement}
\end{align}

For notation convenience, in the rest of subsection, we restate the received time domain signal after CP-removal at the $m$-th SS burst $\mathbf{y}_m\in\mathbb{C}^{P}$ according to the model in Section~\ref{sec:system_model},
\begin{align}
\begin{split}
\mathbf{y}_m =\underbrace{\sum_{l=1}^{L} \tilde{g}_{m,l}\mathbf{Q}(\CFO)\mathbf{F}^{\hermitian}\left[\mathbf{f}(\tau_l)\circ \mathbf{s}\right]}_{\mathbf{x}_m(\boldsymbol{\xi})}+\mathbf{z}_m,
\end{split}
\label{eq:received_signal_for_est}
\end{align}
In the above equation, deterministic vector $\mathbf{x}_m(\boldsymbol{\xi})\in\mathbb{C}^{P}$ is observations model of unknown parameters $\boldsymbol{\xi} \triangleq [\epsilon_{\text{F}},\cdots,\theta_l,\phi_l,\tau_l,\alpha_l,\beta_l,\cdots]^\transpose$, where $\alpha_l = \Re(g_l)$ and $\beta_l = \Im(g_l)$. $\mathbf{z}_m\in\mathbb{C}^{P}$ is the vectorized random noise. We also define $\mathbf{x}(\boldsymbol{\xi}) = [\mathbf{x}^\transpose_1(\boldsymbol{\xi}),\cdots, \mathbf{x}^\transpose_M(\boldsymbol{\xi})]^\transpose.$
Specifically, in (\ref{eq:received_signal_for_est}) vector $\mathbf{s}\in \mathbb{C}^{P}$ contains PSS symbols assigned to $P$ subcarriers. Vector $\mathbf{f}(\tau_l)\in\mathbb{C}^{P}$ is the frequency response
corresponding to the excessive delay $\tau_l$ of a multipath, i.e.,
the contribution of $\tau_l$ on the $p$-th subcarrier is
\begin{align}
[\mathbf{f}(\tau_l)]_p = \mathrm{exp}\left[(-j2\pi (p-1)\tau_l)/(P\Ts)\right].
\label{eq:vector_f}
\end{align} 
Matrix $\mathbf{F}\in\mathbb{C}^{P\times P}$ is discrete Fourier transform (DFT) matrix\footnote{With absence of CFO, multiple DFT matrix $\mathbf{F}$ in $\mathbf{y}_m$ gives frequency domain symbols $\sum_{l=1}^{L}\tilde{g}_{m,l}(\mathbf{f}(\tau_l) \circ \mathbf{s}) + \mathbf{z}_m$.}. The effective channel gain is defined as $\tilde{g}_{m,l} = e^{j\CFO\Nb(m-1)}g_l\mathbf{w}_m^{\hermitian}\mathbf{a}_{\rx}(\phi_l)\mathbf{a}^{\hermitian}_{\tx}(\theta_l)\mathbf{v}_m$, and it includes the contribution of phase rotation across SS bursts due to CFO and IA beamformers $\mathbf{v}_m$ and $\mathbf{w}_m$. Matrix $\mathbf{Q}(\CFO) = \mathrm{diag}\left(\left[1,e^{j\CFO},\cdots,e^{j(P-1)\CFO}\right]^{\transpose}\right)$ contains phase rotations within an OFDM symbol.

%
%
\subsection{Baseline CS formulation}
Directly estimating $\boldsymbol{\xi}$ from (\ref{eq:received_signal_for_est}) via maximum likelihood (ML) requires multi-dimensional search with prohibitive complexity. In the following subsections, we re-formulate $\textit{Problem 2}$ to facilitate sequential parameter estimation.
With straightforward extension of the derivation in \cite[Sec. V]{7400949}, the vector $[\tilde{\mathbf{g}}_l]_m = \tilde{g}_{m,l}$ in (\ref{eq:received_signal_for_est}) can be re-formulated as 
\begin{align}
\tilde{\mathbf{g}}_l = \tilde{\mathbf{Q}}(\CFO)\tilde{\mathbf{A}}^{\hermitian}\mathrm{vec}(\tilde{\mathbf{H}}_l),
\label{eq:CS_formulation}
\end{align}
where $\tilde{\mathbf{A}}\in \mathbb{C}^{G_{\tx}G_{\rx} \times M}$ is defined by the Hermitian conjugate of its $m$-th column as $([\tilde{\mathbf{A}}]_m)^{\hermitian}= (\mathbf{v}_m^{\transpose}\otimes \mathbf{w}_m^{\hermitian} )(\mathbf{A}^{*}_{\tx}\otimes\mathbf{A}_{\rx})$. Note that the above equation is different from \cite[Sec. V]{7400949} which requires $M^2$ sounding beam pairs. The matrix $\tilde{\mathbf{Q}}(\CFO)\in\mathbb{C}^{M\times M}$ contains the phase rotation in each SS burst due to CFO. 
\begin{align}
\tilde{\mathbf{Q}}(\CFO) = \mathrm{diag}\left(\left[1,e^{jN_{\text{B}}\CFO},\cdots,e^{jN_{\text{B}}(M-1)\CFO}\right]^{\transpose}\right).
\label{eq:Q_tilde_matrix}
\end{align}
In fact, matrices $\mathbf{A}_{\tx} \in \mathbb{C}^{N_{\tx} \times G_{\tx}}$ and $\mathbf{A}_{\rx}\in \mathbb{C}^{N_{\rx} \times G_{\rx}}$ are the dictionaries of angular responses with AoAs and AoDs from grids with $\Gt$ and $\Gr$ uniform steps from $-\pi/2$ to $\pi/2$, respectively. In order words, the $k$-th columns in $\mathbf{A}_{\tx}$ and $\mathbf{A}_{\rx}$ are $[\vA{\rx}]_k = \va{\rx}([\mathbf{r}]_k)$ and $[\vA{\tx}]_k = \va{\tx}([\mathbf{t}]_k)$, respectively, where $\{\mathbf{r}\}_k$ are the vectors that contain angle candidates.
\begin{align}
[\mathbf{r}]_k = -\frac{\pi}{2}+(k-1)\Delta\phi, \quad [\mathbf{t}]_k = -\frac{\pi}{2}+(k-1)\Delta\theta.
\label{eq:angle_dictionary}
\end{align}
Also note that the steps $\Delta\theta$ and $\Delta\phi$ depend on the desired resolution. In this work, $G_{\tx}$ and $G_{\rx}$ are used as number of steps and namely $\Delta\theta = 2\pi/G_{\tx}$ and $\Delta\phi = 2\pi/G_{\rx}$. 
Matrix $\tilde{\mathbf{H}}_l \in \mathbb{C}^{\Gr \times \Gt}$ contains the complex path gain of the $l$-th path, i.e., it has $1$ non-zero element whose location depends on the AoA and AoD of the $l$-th cluster in the angular grids.

\textit{Remark 3:} Assuming noisy observation of $\tilde{\mathbf{g}}_l$ and zero CFO, (\ref{eq:CS_formulation}) reduces to the baseline problem in \cite[Sec. V]{7400949}. However, (\ref{eq:received_signal_for_est}) implies that the former assumption is non-trivial unless $\mathbf{s}=\mathbf{1}, \tau_l=0,\forall l$, e.g., \cite{7178503}. Moreover, algorithm designed with latter assumption is sensitive to CFO \cite{hyan_mmWave_CFO}. Finally, the AoA/AoD estimators are commonly confined in $\mathbf{r}$ and $\mathbf{t}$ \cite{7178503}. We address these challenges in the following three subsections.

\subsection{Effective gain estimation}
To address the challenge discussed in \textit{Remark 3}, we propose the following approach. We treat
 $\mathbf{Q}(\CFO)$ in (\ref{eq:received_signal_for_est}) as identity matrix and estimate delay of dominant path and gain, say $\tau_l$ and $\tilde{g}_{m,l}$, by ML approach. Actually, the proposed algorithm uses sparse impulse support $[\mathbf{d}
]_q = q\Delta\tau$ to construct a dictionary, where $\Delta \tau = \Nc\Ts/\Gd$ is the step-size of delay candidates. Based on the knowledge of the model (\ref{eq:vector_f}) and PSS signal $\mathbf{s}$, the delay estimation is implemented as 
\begin{align}
\begin{split}
\hat{q} = & \argmax_{1\leq q \leq G_{\text{D}}}\langle \mathbf{p}_q,\bar{\mathbf{y}}\rangle /\|\mathbf{p}_q\|^2\text{ and }\hat{\tau} = [\mathbf{d}]_{\hat{q}},
\end{split}
\label{eq:delay_est_start}
\end{align}
where $\bar{\mathbf{y}} = \sum_{m=1}^{M}\mathbf{y}_m/M$ is the received PSS samples averaged over $M$ SS bursts. The vector $\mathbf{p}_q$ contains PSS samples when the true delay of dominant path is $[\mathbf{d}]_q$, i.e.,
\begin{align}
\mathbf{p}_q \triangleq  \mathbf{F}^{\hermitian}\left[\mathbf{f}\left([\mathbf{d}]_q\right)\circ \mathbf{s}\right],
\label{eq:delay_vec_dictionary}
\end{align}
where $\mathbf{f}([\mathbf{d}]_q)$ is by plugging in $[\mathbf{d}]_q$ into (\ref{eq:vector_f}). The estimated delay tap $\hat{\tau}$ enables estimating effective gain of a significant path by  
\begin{align}
\hat{\mathbf{g}} = \left(\mathbf{p}_{\hat{q}}^{\hermitian} \otimes \mathbf{I}_M \right)\mathbf{y},
\label{eq:adjust_from_delay_est}
\end{align}
where $\mathbf{I}_M$ is the $M\times M$ identity matrix.

\subsection{Joint AoA and AoD estimation robust to CFO}

The second step uses a modified matching pursuit to solve CS problem (\ref{eq:CS_formulation}) from $\hat{\mathbf{g}}$ while incorporating the existence of CFO in $\tilde{\mathbf{Q}}$. In the conventional matching pursuit step, say the $k$-th, the anticipated effective channel response corresponding to an AoA and AoD pair, i.e., $[\tilde{\mathbf{A}}]_k$, is used to evaluate inner product with $\mathbf{g}$ \cite{7390019}. The proposed heuristic treats AoA and AoD as known in the $k$-th step, and uses the ML estimator of CFO $\hat{\epsilon}_{\text{F},k}$ which is available in closed form. The modified matching pursuit is expressed as
\begin{align}
\begin{split}
\hat{k} = \argmax_{1\leq k \leq G_{\text{R}}G_{\text{T}}}\langle \tilde{\mathbf{Q}}(\hat{\epsilon}_{\text{F},k})\tilde{\mathbf{a}}_k,\hat{\mathbf{g}}\rangle /\|\tilde{\mathbf{a}}_k\|^2,
\end{split}
\label{eq:AoA_AoD_est_start}
\end{align}
where $\tilde{\mathbf{a}}_k \triangleq [\tilde{\mathbf{A}}]_k$ from (\ref{eq:CS_formulation}). The matrix $\tilde{\mathbf{Q}}(\hat{\epsilon}_{\text{F},k})$ has structure as (\ref{eq:Q_tilde_matrix}). The input $\hat{\epsilon}_{\text{F},k}$ is the ML CFO estimator when treating AoA/AoD as they correspond to ones in $[\tilde{\mathbf{A}}]_k$. Specifically, the CFO estimator relies on the estimator in \cite{Kay_single_tone} by treating $\bar{\mathbf{y}}_{k} = \tilde{\mathbf{a}}^{*}_k \circ \hat{\mathbf{g}}$ as a tone with frequency $\CFO$. 
\begin{align}
\begin{split}
\hat{\epsilon}_{\text{F},k} = &\frac{1}{N_{\text{B}}}\angle\left(\frac{1}{M-1}\sum_{m=1}^{M-1}\left[\bar{\mathbf{y}}_k\right]^{*}_{m}\left[\bar{\mathbf{y}}_k\right]_{m+1}\right).
\end{split}
\label{eq:CFO_est}
\end{align} 
Operation $\angle(x) = \tan^{-1}[\Im(x)/\Re(x)]$ evaluates angle based on complex samples. 
To get estimates of the AoA, AoD, and CFO, index $\hat{k}$ is used to select candidates from grids (\ref{eq:angle_dictionary}) after the following adjustment $\hat{k}_{\rx} = \lfloor (\hat{k}-1)/\Gt \rfloor+1$ and $\hat{k}_{\tx} = \hat{k} - (\hat{k}_{\rx}-1)\Gt$,
\begin{align}
\hat{\phi} = [\mathbf{r}]_{\hat{k}_{\rx}}, \quad \hat{\theta} = [\mathbf{t}]_{\hat{k}_{\tx}},\quad \hat{\epsilon}_{\text{F}} = \hat{\epsilon}_{\text{F},\hat{k}}.
\label{eq:get_est_from_dictionary}
\end{align}

\subsection{Off-grid refinement} 

The aformentioned heuristics provide estimates of delay, AoA, and AoD that are restricted to the grid, i.e., elements of $\mathbf{d},\mathbf{r}$ and $\mathbf{t}$. Grid refinement is a technique to provide off-grid estimation accuracy. There are several approaches considered in the literature including multi-resolution refinement \cite{1468495} and the Newtonized gradient refinement \cite{7491265}. In this work, we propose to use first order descent approach. As initialization of refinement, the estimator from previous steps is saved into $\hat{\boldsymbol{\xi}}^{(k)}$ for $k=1$. In the $k$-th iteration, the following error vector is evaluated
\begin{align}
\mathbf{e}^{(k)} = \mathbf{y} - \mathbf{x}\left(\hat{\boldsymbol{\xi}}^{(k)}\right),
\label{eq:residual_error}
\end{align}
where $\mathbf{y}$ is the received signal after rearrangement as (\ref{eq:rearrangement}), $\mathbf{x}(\hat{\boldsymbol{\xi}}^{(k)})$ is obtained by plugging in estimated parameters into parametric model (\ref{eq:received_signal_for_est}). In other words, $\mathbf{e}^{(k)}$ is the error vector between observed signal sequence and received signal model using current estimates, which is then used to update parameters. The complex gain in iteration $k$ is computed as
\begin{align}
\hat{g}^{(k+1)} = (\boldsymbol{\nabla}\mathbf{x}_{g})^{\dagger}\mathbf{y},
\label{eq:gain_refinement}
\end{align}
where $\boldsymbol{\nabla}\mathbf{x}_{g} = (\partial \mathbf{x}(\boldsymbol{\xi})/\partial g)|_{\boldsymbol{\xi} = \hat{\boldsymbol{\xi}}^{(k)}}$ is the partial derivative of $\mathbf{x}(\boldsymbol{\xi})$ over parameter $g$ in (\ref{eq:received_signal_for_est}) evaluated at $\hat{\boldsymbol{\xi}}^{(k)}$. The refinement steps for delay, CFO, AoA, and AoD are moving towards the gradient of their estimators in the previous iterations. For concise notation, in the following equation and paragraph we use $x$ to denote the parameter to be refined, i.e., $x=\{\tau, \CFO,\theta,\phi\}$. The refinement steps are

\begin{align}
\begin{split}
\hat{x}^{(k+1)} = \hat{x}^{(k)}+ \mu_{x}\Re\left[(\boldsymbol{\nabla}\mathbf{x}_{x})^{\dagger}\mathbf{e}^{(k)}\right], x=\{\tau, \CFO,\theta,\phi\},
\end{split}
\label{eq:refinement}
\end{align}
where $\mu_{x}$ is the step-size, vector $\boldsymbol{\nabla}\mathbf{x}_{x} = (\partial \mathbf{x}(\boldsymbol{\xi})/\partial x)|_{\boldsymbol{\xi} = \hat{\boldsymbol{\xi}}^{(k)}}$ is the the partial derivative of $\mathbf{x}(\boldsymbol{\xi})$ in (\ref{eq:received_signal_for_est}) over parameter of interest. The above approach iteratively runs by appending updated parameter into $\mathbf{x}(\hat{\boldsymbol{\xi}}^{(k+1)})$ for the next iteration until the error $\|\mathbf{e}^{(k)}\|^2$
converges or falls below threshold $\epsilon_0$.

It is worth noting that the proposed approach can be extended to support multi-path training which has been covered by a variety of works in CS-based approaches \cite{7390019,8306126,8356247,8488662,8323164,Myers_CFO_SPAWC_1707,8309152}. However, the main motivation of this work is to showcase and analyze pseudorandom sounding beams in the initial access and initial beam training. Thus the only metric directly comparable to its counterparts \cite{7161389,Caire_DIA_arxiv_1709,UTA_stocha_geo_IA_TCOM_18,KTH_stoch_geo_IA_arxiv_1802,Meng_OIA_TCOM_1803,Zorzi_IA_MCOM_1611,UTA_IA_TWC_1705,Giordani_beam_turotial_arxiv_1804}, namely single path training, is evaluated. 

The entire compressive initial beam training algorithm is summarized in Algorithm 1.
\begin{algorithm}
\caption{Compressive Initial Access and BF Training}
\begin{algorithmic}[1]
\Statex \textbf{Input:} Received IA signal sequence $y[n]$
\Statex \textbf{Output:} Discovery decision; Beam pair $\mathbf{v}^{\star}, \mathbf{w}^{\star}$ 
  \Statex $\mathrm{\%}$\quad\quad\quad--------- \quad  Initial Discovery \quad ---------\quad\quad
  \State PSS correlation (\ref{eq:ZC_correlation}).
  \State Energy detection (\ref{eq:ED_w_STO}) and timing acquisition (\ref{eq:timing_acquisition}).
  \If{$\mathrm{Positive Decision}$}
  \Statex $\mathrm{\%}$\quad\quad\quad------ \quad  Initial BF Training (Coarse) \quad ------\quad\quad
  \State Arrange sequence $y[n]$ into vector $\mathbf{y}$ as (\ref{eq:rearrangement}).
  \State{Estimate excessive delay as (\ref{eq:delay_est_start})}.
  \State{Estimate effective channel gain as (\ref{eq:adjust_from_delay_est}).}
  \State Matching pursuit (\ref{eq:AoA_AoD_est_start}) with CFO estimation (\ref{eq:CFO_est}).
  \State Get AoA, AoD, and CFO estimators in (\ref{eq:get_est_from_dictionary}).
  \Statex $\mathrm{\%}$\quad\quad------ \quad  Initial BF Training (Fine) \quad ------\quad
  \While{$\|\mathbf{e}^{(k)}\| > \epsilon_0$ in (\ref{eq:residual_error})} 
  \State{Use refinement steps (\ref{eq:gain_refinement}) and (\ref{eq:refinement}); $k=k+1$.}
  \EndWhile\label{euclidendwhile}
  \State{ Report beam pair $\mathbf{w}^{\star} = \mathbf{a}_{\rx}(\phi^{(k)}_l)$, $\mathbf{v}^{\star} = \mathbf{a}_{\tx}(\theta_l^{(k)})$.}
  \EndIf
\end{algorithmic}
\end{algorithm}

%
%
\subsection{Performance bound of initial BF training in LOS}
In this subsection, we provide lower bound of AoA/AoD estimation variance in pure LOS scenario, namely CRLB in joint estimating $\boldsymbol{\xi} = [\epsilon_{\text{F}},\theta_1,\phi_1,\tau_1,\alpha_1,\beta_1]^\transpose$.
%
Based on (\ref{eq:received_signal_for_est}), the likelihood function is 
$\prob(\mathbf{y};\boldsymbol{\xi}) = (2\pi\sigman^{2MP})^{-1}\text{exp}\left(-(\|\mathbf{y}-\mathbf{x}(\boldsymbol{\xi})\|^2)/(\sigman^2)\right)$.
The log-likelihood function is $L(\mathbf{y};\boldsymbol{\xi}) \triangleq \ln [\prob(\mathbf{y};\boldsymbol{\xi})]$. The lower bound of estimation variance is given in the following proposition.

\begin{proposition}
\label{proposition:CRLB}
 The CRLB of AoA/AoD estimation in the compressive initial BF training stage in LOS environment is 
 \begin{align}
 \begin{split}
\mathrm{var}(\hat{\phi}_1) \geq 
 [\mathbf{J}^{-1}]_{2,2},\quad
 \mathrm{var}(\hat{\theta}_1) \geq
 [\mathbf{J}^{-1}]_{3,3}
 \end{split}
 \end{align}
 where $\mathbf{J}\triangleq \partial^2 L(\mathbf{y};\boldsymbol{\xi})/\partial\boldsymbol{\xi}^2$ is the Fisher Information Matrix (FIM) whose expressions are listed in Appendix~\ref{app:CRLB}.
\end{proposition}
\begin{IEEEproof}
See Appendix~\ref{app:CRLB}.
\end{IEEEproof}
\subsection{Benchmark approach: hierarchical directional BF training}
The directional beams in SS burst allow BS and UE to coarsely estimate the propagation directions \cite{8323183}. Although approach in \cite{8323183} is not tailored for wideband channel with synchronization offset, it relies on RSS measurement within burst and therefore it is robust to the model mismatch. Using the SS burst index that corresponds to the maximum received power, the system uses the knowledge of directional sounding beams to infer channel propagation angles. Specifically, as illustrated in Fig.~\ref{fig:frame_structure}, the estimated $\theta^{\star}$ and $\phi^{\star}$ are the centers of the $\hat{m}_{\tx}$-th and $\hat{m}_{\rx}$-th sounding beams in BS and UE \cite{8323183}, respectively.  Note that the estimated angle sector indices $\hat{m}_{\tx}$ and $\hat{m}_{\rx}$ are computed from the SS burst index $m_{\text{DIA}}^{\star}$ in (\ref{eq:DIA}), i.e., $\hat{m}_{\rx} = \lfloor (m_{\text{DIA}}^{\star}-1)/M_{\tx} \rfloor+1$, and $\hat{m}_{\tx} = m_{\text{DIA}}^{\star}-(\hat{m}_{\rx}-1)M_{\tx}$. 
The large width of a sector beam results in poor angular resolution in DIA. In order to improve the resolution, hierarchical directional beam training scans narrower beams within the sector of interest. Such procedure occurs during CSI-RS bursts which are scheduled for individual UEs. 

\section{Access latency, overhead and DSP complexity}
\label{sec:comparison_analysis}

In this section, we present a model for analyzing three system performance indicators, namely access latency, overhead, and computational complexity. Note that this unified model applies to both directional scheme and the proposed approach. 

\begin{figure}
\begin{center}
\includegraphics[width=0.5\textwidth]{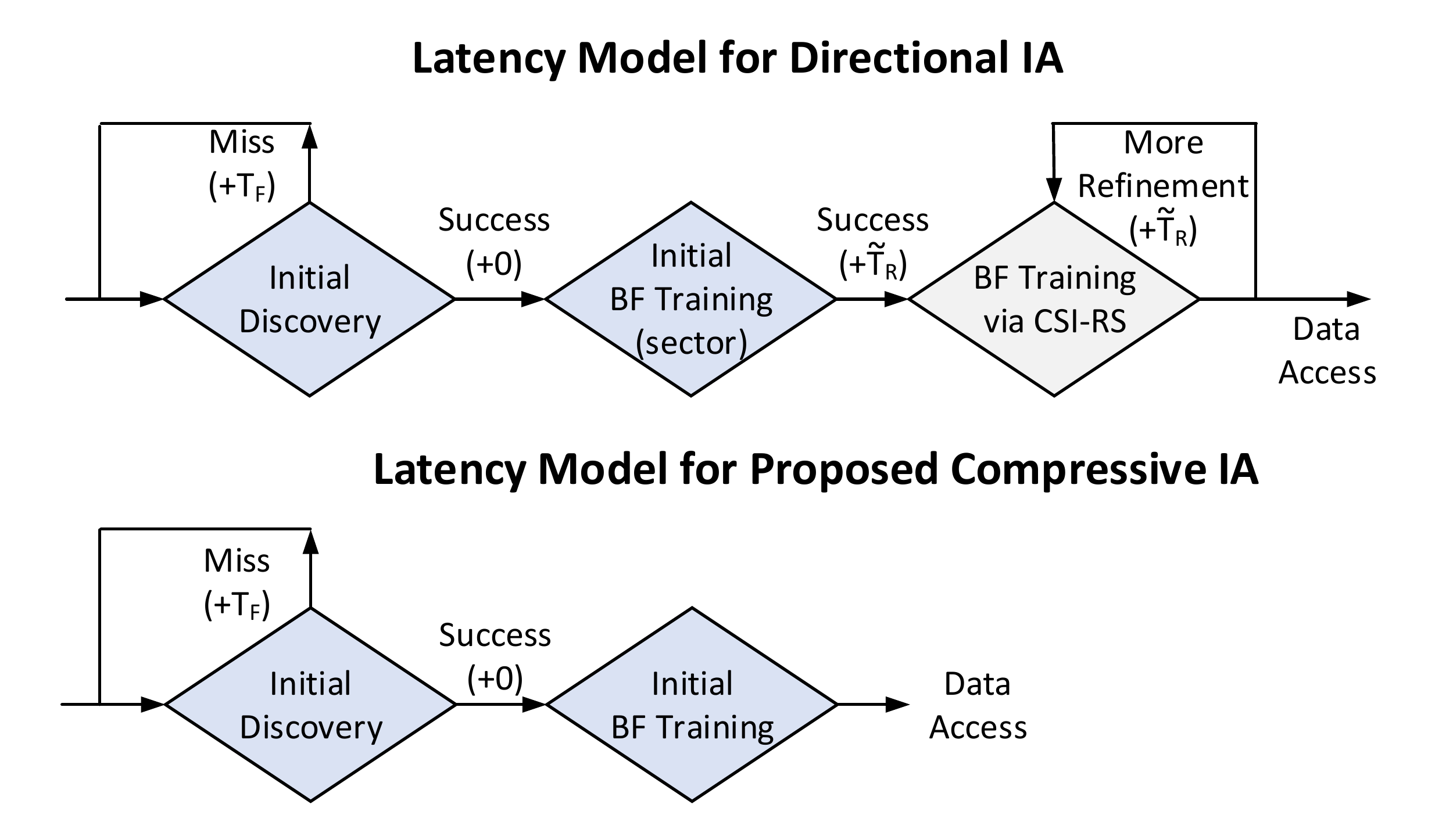}
\end{center}
\vspace{-4mm}
\caption{Initial access latency model for directional IA and the proposed compressive IA. The associated latency in each steps are shown in the bracket.}
\vspace{-4mm}
\label{fig:latency_model}
\end{figure}

Based on \cite{Giordani_beam_turotial_arxiv_1804}, we propose to use the latency model for both SS burst and CSI-RS as shown in Fig.~\ref{fig:latency_model}. In both IA schemes, the failure of cell discovery introduces penalty of $\TSS$ for a new IA block. When cell discovery occurs, the additional latency is required for scheduled CSI-RS according to the required number $N_{\text{train}}$. Thus the access latency is
\begin{align}
T_{\text{latency}} = \sum_{k=0}^{+\infty}P^{k}_{\text{MD}}(1-P_{\text{MD}})k\TSS + \tilde{T}_{\text{R}}N_{\text{Train}}
\label{eq:latency}
\end{align}
where the first term includes latency for cell discovery. In the second term, $\tilde{T}_{\text{R}}$ is the average time for the UE to get the scheduled CSI-RS for beam training and it is expressed as
\begin{align*}
\tilde{T}_{\text{R}} = \frac{1}{\NUE}\left[\sum_{k=0}^{K_{\text{F}}}\sum_{q=1}^{K_{\text{R}}}((k-1)\TSS+qT_{\text{R}})+ \sum_{q=1}^{K_{\text{res}}}(K_{\text{F}}\TSS+qT_{\text{R}})\right]
\end{align*}
In the above equation, $\NUE$ denotes the number of UEs in the network. They share available CSI-RS in a time division manner to combat channel dynamic. Due to the limited number of CSI-RS $K_{\text{R}} = \lfloor(\TSS-MT_{\text{B}})/T_{\text{R}}\rfloor$ within one IA period, more than one frame duration is required to meet scheduling of large number of UE $\NUE$. Therefore, in (\ref{eq:latency}) $K_{\text{F}} = \lfloor (\NUE-1)/K_{\text{R}}\rfloor$ is the number of frames required to assign all CSI-RS to UEs and $K_{\text{res}} = \NUE - K_{\text{cyc}}K_{\text{R}}$ is the residual delay in the last frame. As shown in the next section, DIA and directional BF training typically require larger $N_{\text{train}}$ than the proposed approach.



%
%
Following \cite{Giordani_beam_turotial_arxiv_1804}, the overhead (OH) ratio is modeled by counting the time-frequency resource in IA and CSI-RS 
\begin{align}
\text{OH} = \frac{MB_{\text{IA}}\Tb + K_{\text{R}}B_{\text{tot}}T_{\text{r}}}{B_{\text{tot}}\TSS} \times 100\%
\label{eq:overhead}
\end{align}
where $B_{\text{IA}} = 1/T_{\text{s}}$ is the bandwidth in IA and the channel usage is $M\Ts$ every period $\TSS$. We focus on varying CSI-RS density $K_{\text{R}}$. Note that with reduced $K_{\text{R}}$ (increased $T_{\text{R}}$), the OH reduces
with a cost of additional latency. 



The required baseband operations of the proposed approach are summarized in Table~\ref{tab:operations}, where only the complex multiplications are taken into account.
As shown in later sections, the off-grid refinement in the proposed algorithm provides CRLB reaching accuracy but is necessary to reach adequate beam alignment. As a consequence, we do not include its complexity here. It is worth noting that the above analysis assumes there is an off-line pre-computation of all required dictionaries for matching pursuit, i.e., $\mathbf{p}_q$ in (\ref{eq:delay_est_start}), $\tilde{\mathbf{a}}_k$ in (\ref{eq:AoA_AoD_est_start}). In addition, the directional IA requires the first two steps in Table~\ref{tab:operations}. 

\begin{table}
\caption{Digital Baseband Operations (complex multiplications)}
\centering
\begin{tabular}{|c|c|c|}
\hline 
\textbf{Function Block} & \textbf{Equation}&\textbf{Operations}\tabularnewline
\hline 
\hline 
\multicolumn{3}{|c|}{Initial Discovery} \tabularnewline
\hline
PSS FIR corr. & (\ref{eq:ZC_correlation}) & $P\Nb$ \tabularnewline
\hline
Detection and time sync. & (\ref{eq:ED_wo_STO}) or (\ref{eq:ED_w_STO}) & $\Nb$ \tabularnewline
\hline
\multicolumn{3}{|c|}{Initial BF training} \tabularnewline
\hline
Excess. delay est. & (\ref{eq:delay_est_start}) and (\ref{eq:adjust_from_delay_est}) & $P\Gd +PM$ \tabularnewline
\hline
AoA/AoD est. & (\ref{eq:AoA_AoD_est_start}) & $M\Gt\Gr$ \tabularnewline
\hline
CFO est. & (\ref{eq:CFO_est}) & $2M\Gt\Gr$ \tabularnewline
\hline
\end{tabular}
\vspace{-1mm}
\label{tab:operations}
\end{table}

%
%
\section{Results}
\label{sec:simulation results}
This section presents the numerical comparison between the proposed approach and DIA with directional beam training.

%
%
\subsection{Simulation settings}

\begin{figure}
\begin{center}
\includegraphics[width=0.45\textwidth]{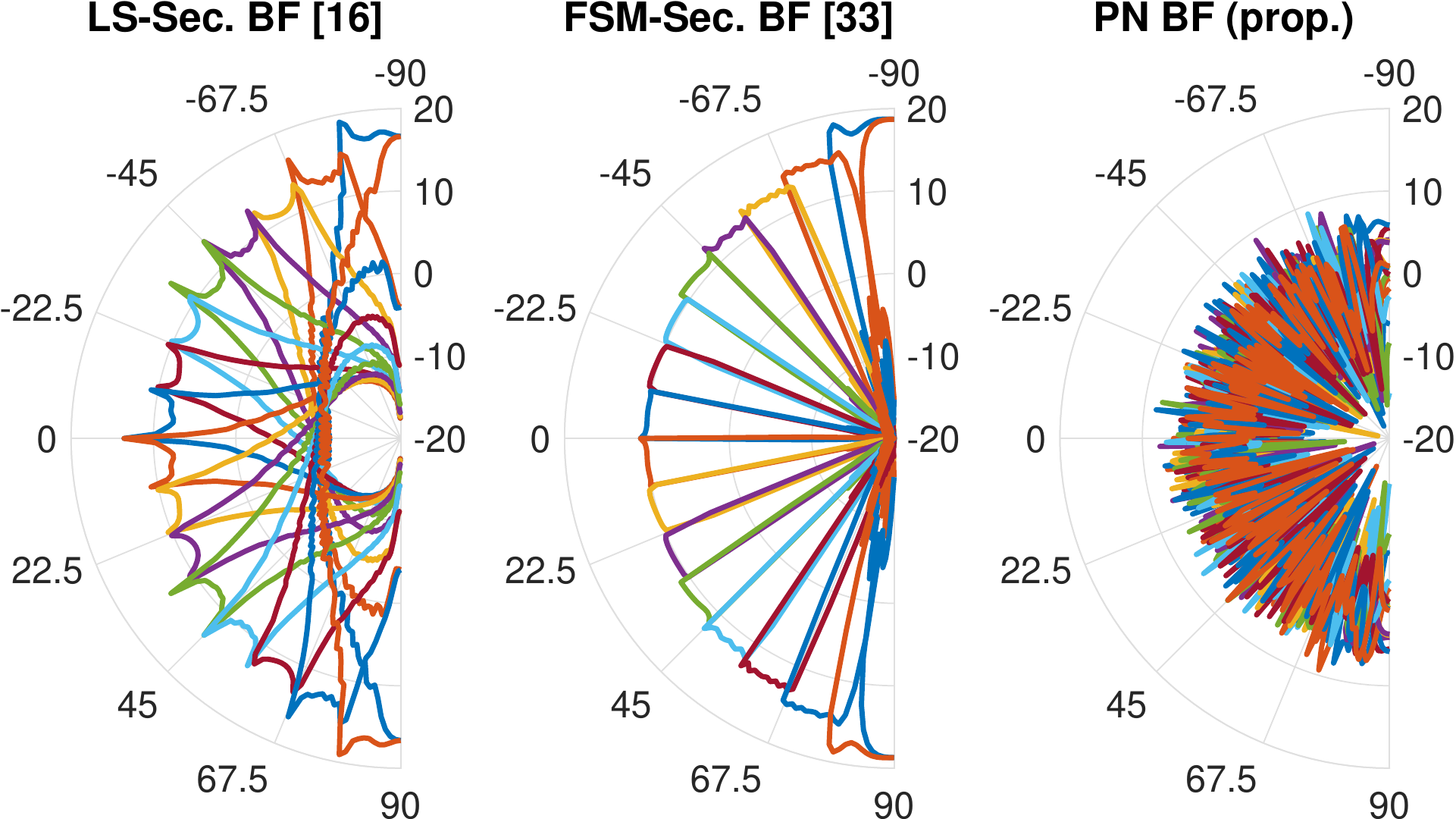}
\end{center}
\vspace{-4mm}
\caption{Beam patterns of two sector beam designs \cite{6847111,array_textbook} with $M_{\text{T}}=16$ transmit sectors and one realization of 16 pseudorandom beams. In the polar plot, the $r$-axis refers to the gain in dB and the angular axis refers to steering angle in degrees. All patterns correspond to $N_{\tx} = 128$ ULA.}
\vspace{-4mm}
\label{fig:beam_pattern}
\end{figure}

The simulations follow 5G-NR frame structure. We first evaluate performance in the simplified 2D S-V channel model. The maximum excess delay is set as $\Nc=4$ samples. As for the benchmark DIA, we use two approaches to design directional sector beams, i.e., least-squares based sector beamforming (LS-Sec) codebook \cite{6847111} and frequency sampling method based sector beamforming (FSM-Sec) codebook \cite[C23.4]{array_textbook}. Examples of beam patterns\footnote{We uses an optimistic benchmark system where sector beams are synthesized by arrays with ideal phase and magnitude control.} are shown in Fig.~\ref{fig:beam_pattern}. In each of the Monte Carlo simulations, we generate an independent random realization of pseudorandom sounding beam codebook and channel parameters, unless otherwise mentioned. 

We also evaluate the efficacy of the proposed approach in a realistic 3D mmW environment. In Section~\ref{sec:realistic_sim} we simulate the system with QuaDRiGa simulator \cite{6758357} based on mmMAGIC model \cite{mmMAGIC_model} in $28$ GHz urban-micro (UMi). We remove \textit{Assumptions 1, 2} from Sections~\ref{sec:signal_model_before_timing} and \ref{sec:signal_rearrangement}. Uniform planar arrays (UPA) $N_{\tx} = 16 \times 4$, $N_{\rx} = 4 \times 4$ are used at BS and UE, respectively, to exploit the higher sparsity in the elevation plane. The proposed algorithm is extended to UPA and 3D mmW channel model accordingly. In the simulations, the transmit power is set to $P_{\text{out}} = 46$ dBm. The large scale channel model includes pathloss and shadowing. The AWGN on the receiver with 4dB noise figure is added with power of $-170+10\log_{10}(\text{BW})$ dBm, where $\text{BW}=1/\Ts$ and $\text{BW} = B_{\text{tot}} = 400$ MHz \cite{8316765} for IA and data stage, respectively. Moreover, the UE phase noise, $\psi[n]$ in (\ref{eq:rx_signal_raw}), is modeled as Weiner process \cite{847872} that corresponds to oscillator with phase noise spectrum $-114$ dBc/Hz at $1$ MHz offset 
\cite{7508265}.
The frame structure remains as described previously and other detailed simulations setting in QuaDRiGa can be found in the supplementary material \cite{hyan_git}. The benchmark approaches are also extended for UPA and 3D channel, i.e., FSM-Sec beams are extended in both azimuth and elevation plane. During each one of $N_{\text{train}}$ CSI-RS, BS and UE use 16 sounding beams pairs which bisect previous scanned azimuth and elevation angular regions. We use post-training SNR as performance indicator, which is evaluated by dividing channel gain $P_{\text{out}}\sum_{d=0}^{N_\text{c}}|(\mathbf{w}^{\star})^{\hermitian}\mathbf{H}[d]\mathbf{v}^{\star}|^2$ over noise power in $B_{\text{tot}}$. 



Unless otherwise mentioned, the simulation parameters are summarized in Table.~\ref{tab:simulation_setting}.

\begin{table}
\caption{Summary of Simulation settings}
\centering
\begin{tabular}{|l|l|}
\hline 
\textbf{Parameters} & \textbf{Values in Simulations}\tabularnewline
\hline 
\hline 
\multicolumn{2}{|c|}{Frame Structure} \tabularnewline
\hline
SS Signal BW & $1/\Ts = 57.6$ MHz \cite{Giordani_beam_turotial_arxiv_1804}\tabularnewline
\hline 
Carrier and PSS Length & $P = 128$ \cite{Giordani_beam_turotial_arxiv_1804} \tabularnewline
\hline
Max Exces. Delay and CP & $\Nc = \{4,32\}$ and $\Ncp = \{8,32\}$\tabularnewline
\hline 
SS Burst Duration & $\Nb = 1024$ ($\Tb =17.84\mu$s) \cite{Giordani_beam_turotial_arxiv_1804} \tabularnewline
\hline 
SS Burst Num. & $M = 64$\cite{Giordani_beam_turotial_arxiv_1804}; $M_{\text{T}} = 16$, $M_{\text{R}} = 4$ \tabularnewline
\hline
SS Signal Period & $\TSS = 20$ ms \cite{Giordani_beam_turotial_arxiv_1804}\tabularnewline
\hline 
\hline  
\multicolumn{2}{|c|}{Initial Synchronization Offset} \tabularnewline
\hline
Freq. Offset at UE & Up to $\pm$5 ppm \cite{CFO_IA_spec} \tabularnewline
\hline 
Timing Offset at UE& $\epsilon_{\text{T}} = \{170,960\}$, ($\Delta\tau = 3,17\mu$s) \tabularnewline
\hline
STO Search Window& $\epsilon_{\text{T,max}} = 1024$ \tabularnewline
\hline
\hline
\multicolumn{2}{|c|}{Algorithm Design} \tabularnewline
\hline
Target False Alarm& $P^{\star}_{\text{FA}} = 0.01$ \tabularnewline
\hline
Dictionary Size & $\Gd = 500$, $\Gt = 2N_{\tx}$, $\Gr = 2N_{\rx}$ \tabularnewline
\hline
\end{tabular}
\vspace{-1mm}
\label{tab:simulation_setting}
\end{table}

%
%
\subsection{Performance in simplified S-V channel model}
\begin{figure}
\begin{center}
\includegraphics[width=0.52\textwidth]{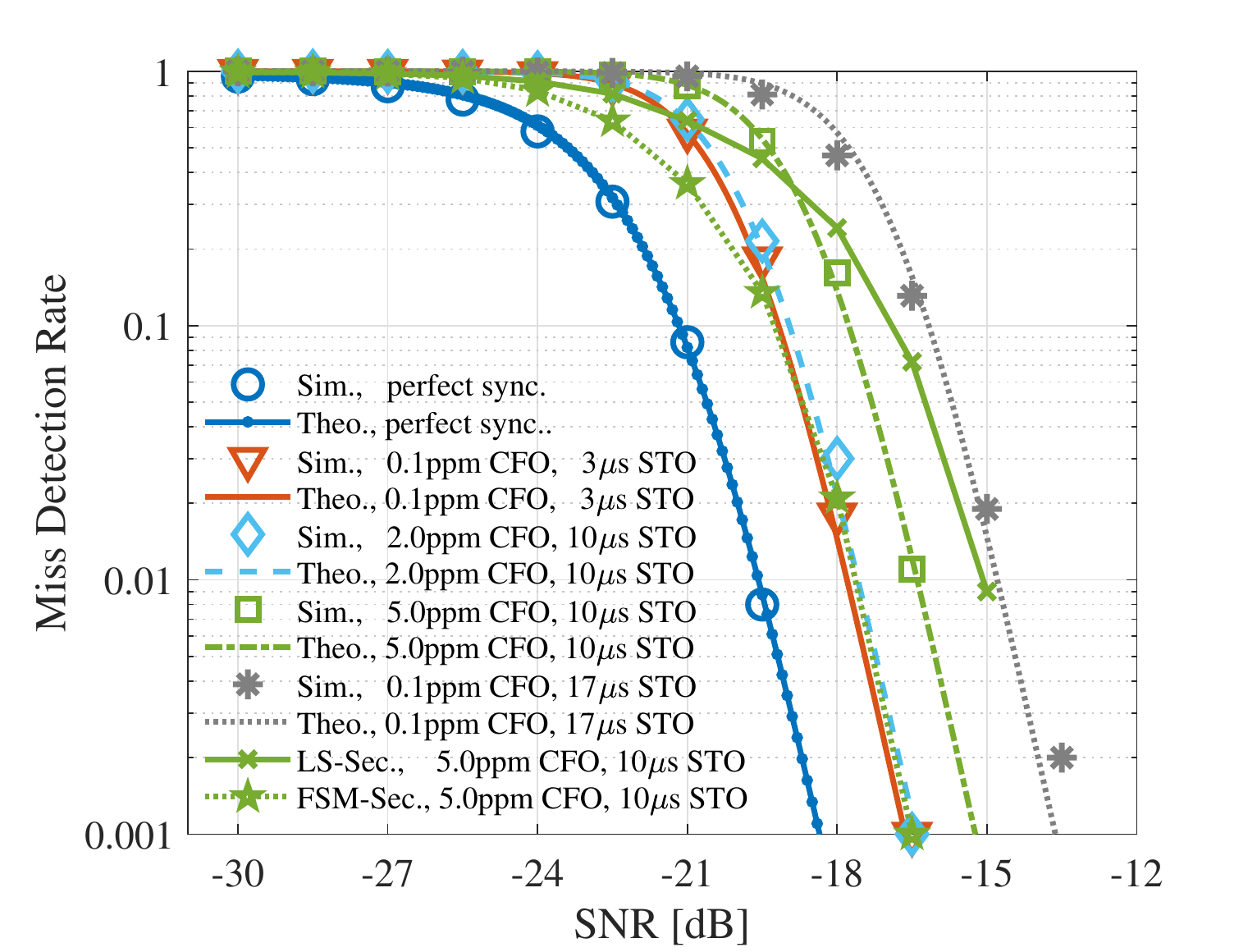}
\end{center}
\vspace{-4mm}
\caption{Simulated (Sim.) and theoretical (Theo.) results of the miss detection rate of the proposed initial discovery with various synchronization errors. The discovery rate of the directional initial access is also included as benchmark and both LS-Sec. and FSM-Sec. are used as sector beams. The BS and UE have $N_{\tx} = 128$ and $N_{\rx}=32$ ULA and SV channel has $L=2$ multipaths.}
\vspace{-4mm}
\label{fig:detection_theo_vs_sim}
\end{figure}

The miss detection rate\footnote{Miss detection rate in simulation is evaluated by a generalized definition $\prob(\gamma_{\text{NT}} >\eta_{\text{NT}}, \hat{\STO} = \STO|\mathcal{H}_1)$ in this proposed approach when $\STO \neq 0$.} of the proposed approach for initial discovery is shown in Fig.~\ref{fig:detection_theo_vs_sim}, and it is verified against the theoretical expressions (\ref{eq:PMD_w_STO}). We have the following findings. Firstly, the lack of perfect timing synchronization introduces around 3 dB sensitivity loss as shown between the blue circled curve and red solid curve. However, this issue is unavoidable in practical systems. Secondly, less than 3 dB sensitivity loss occur when $\pm$5 ppm CFO is present in addition to STO, as shown by the light blue dashed and green dashed-and-dotted curves. Finally, the practical STO ($\leq10\mu$s) is noncritical as shown by red solid and blue dashed curves. But when STO is large enough to cause transmitter and receiver burst beamforming window mismatch, e.g., 17 $\mu$s STO which corresponds to large $K(\STO)$ in (\ref{eq:define_gain_split_K}), severe sensitivity loss is introduced as shown in grey dotted curves. In summary, these simulations verified the findings from Section~\ref{sec:cell_discovery} that practical initial synchronization error introduces up to few dB sensitivity loss as compared to perfect synchronization scenario. 

The comparison among proposed approach and benchmark DIA based discovery approaches is also presented in Fig.~\ref{fig:detection_theo_vs_sim}. Although common sense may doubt the efficacy of the proposed approach since there is no significant angular gain for any beam pattern, as illustrated in Fig.~\ref{fig:beam_pattern}, the results show that there is only a couple of dB difference among the proposed approach and benchmark. However, such gap is less than the performance fluctuation of DIA with difference codebooks. The rationale behind this result is that the proposed scheme collects signal energy spread over all $M$ SS bursts which in fact gives equivalent energy measurement as directional approach where energy collection occurs only when a sector beam aligns with true propagation direction.

%
%

\begin{figure}
\begin{center}
\includegraphics[width=0.48\textwidth]{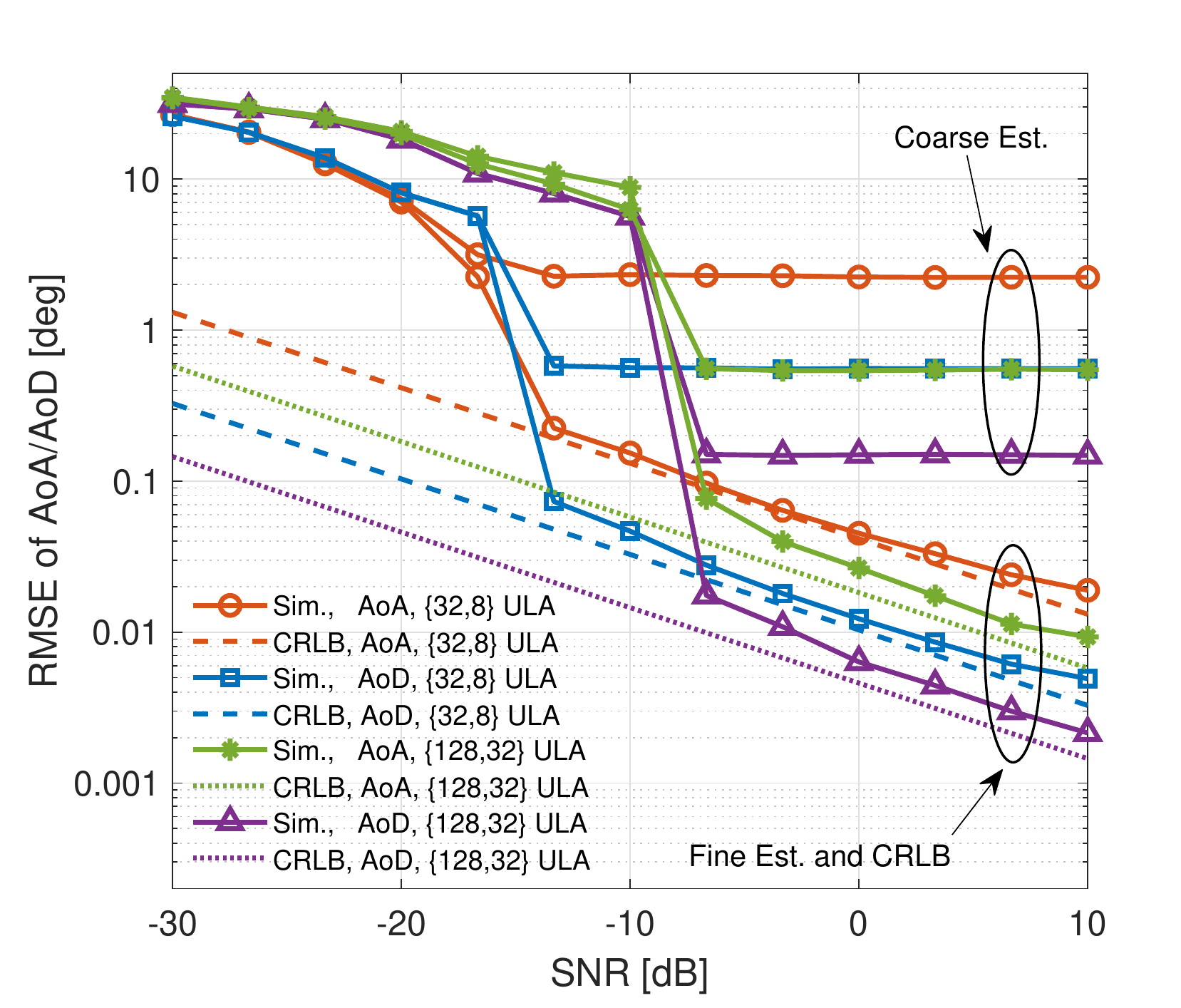}
\end{center}
\vspace{-4mm}
\caption{Simulated results of the proposed algorithm, with and without refinement steps, and theoretical bound of RMSE of AoA/AoD estimation in LOS. Both array geometry setting, $\{N_{\tx},N_{\rx}\} = \{32,8\}$ ULA and $\{N_{\tx},N_{\rx}\} = \{128,32\}$ ULA are evaluated. System has 5ppm CFO.}
\vspace{-4mm}
\label{fig:CRLB}
\end{figure}

The beam training performance of the proposed BF training algorithm in LOS is presented in Fig.~\ref{fig:CRLB}. The performance metrics are the residual mean square error defined by $\text{RMSE}_{\text{AoA}} = \sqrt{\mathbb{E}|\hat{\phi}_1-\phi_1|^2}$ and $\text{RMSE}_{\text{AoD}} = \sqrt{\mathbb{E}|\hat{\theta}_1-\theta_1|^2}$. The simulations are conducted with \textit{Assumption 2}. The same pseudorandom setting is used in both simulation and theoretical CRLB evaluation. The refinement steps are forced to terminate in up to 100 iterations. We have the following findings. Firstly, when the off-grid refinement are used, the proposed algorithm reaches CRLB in high SNR regime. Secondly, the coarse estimation in high SNR has a compromised performance as compared to CRLB. However coarse estimation (without refinement) has adequate accuracy for beam steering since RMSE is order of magnitude lower than $3$ dB beam-width in steering, i.e., $0.29\pi/N_{\tx}$ and $0.29\pi/N_{\rx}$. Finally, Fig.~\ref{fig:detection_theo_vs_sim} and \ref{fig:CRLB} reveal that in SNR region between $-15$ dB and $-7.5$ dB reliable detection occurs but beam training performance is poor. Admittedly, this implies a compromised experience for UEs at the cell edge, which is worth further investigation.

%
%
\subsection{Performance in QuaDRiGa channel simulator}
\label{sec:realistic_sim}

\begin{figure*}[htp]
\begin{tabular}{p{6.7cm}p{12.5cm}}
\subfloat[Network illustration where a UE is randomly distributed in the horizontal plane with height of UE within 20 meters. ]{%
  \includegraphics[clip,width=0.65\columnwidth]{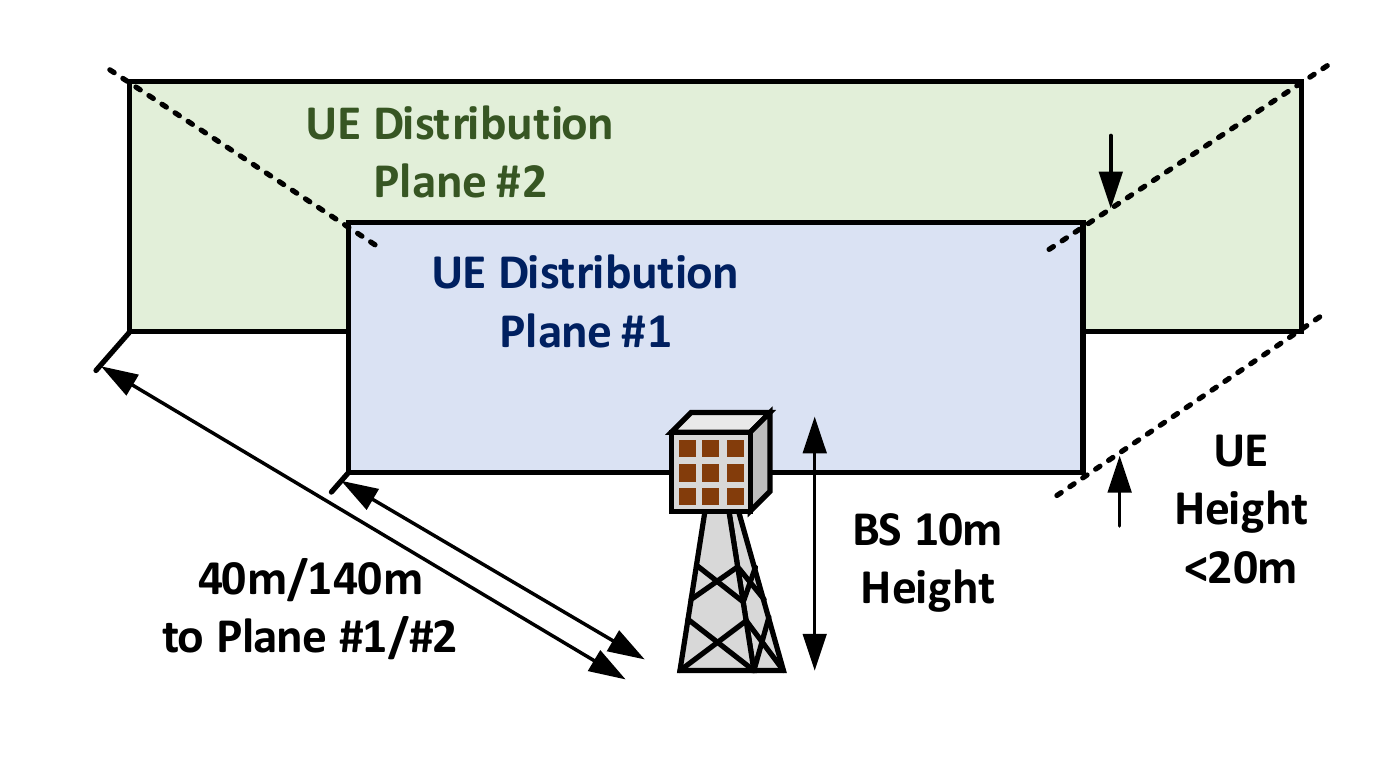}%
} &
\multirow{1}[5]{*}{\subfloat[The CDF of post-training beam steering SNR in the data phase. For the DIA, different number of CSI-RS $N_{\text{train}}$ are considered. The SNR distribution corresponding to beam steering towards true LOS path (when existing) is also included as benchmark. The miss detection rates are included in each plots.]{%
  \includegraphics[clip,width=1.15\columnwidth]{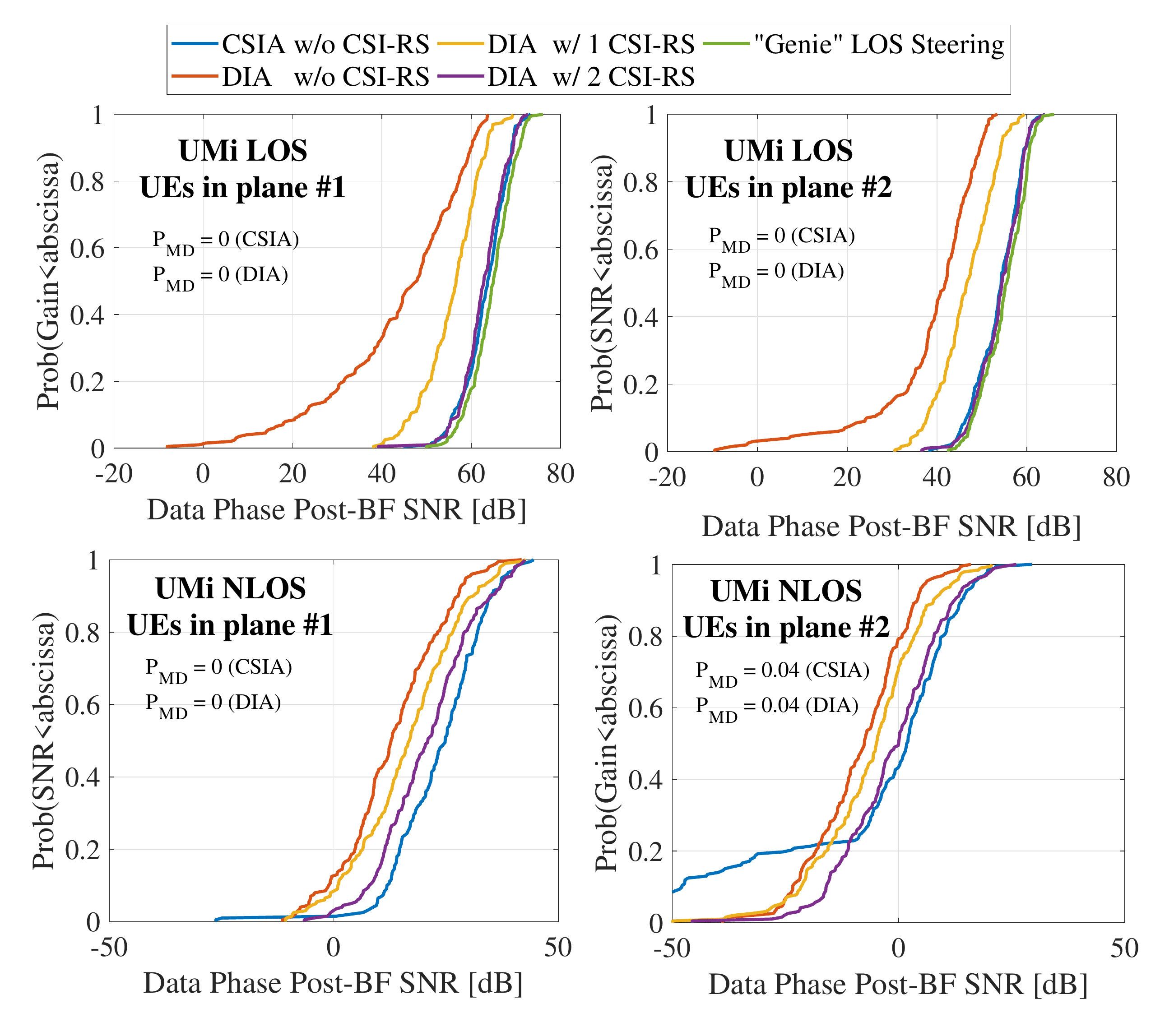}%
}}\\
\subfloat[Access latency versus overhead of both IA scheme  (\ref{eq:latency}) and (\ref{eq:overhead}) with different number of UEs that share the scheduled CSI-RS. $P_{\text{MD}}$ is 0.04.]{%
  \includegraphics[clip,width=0.73\columnwidth]{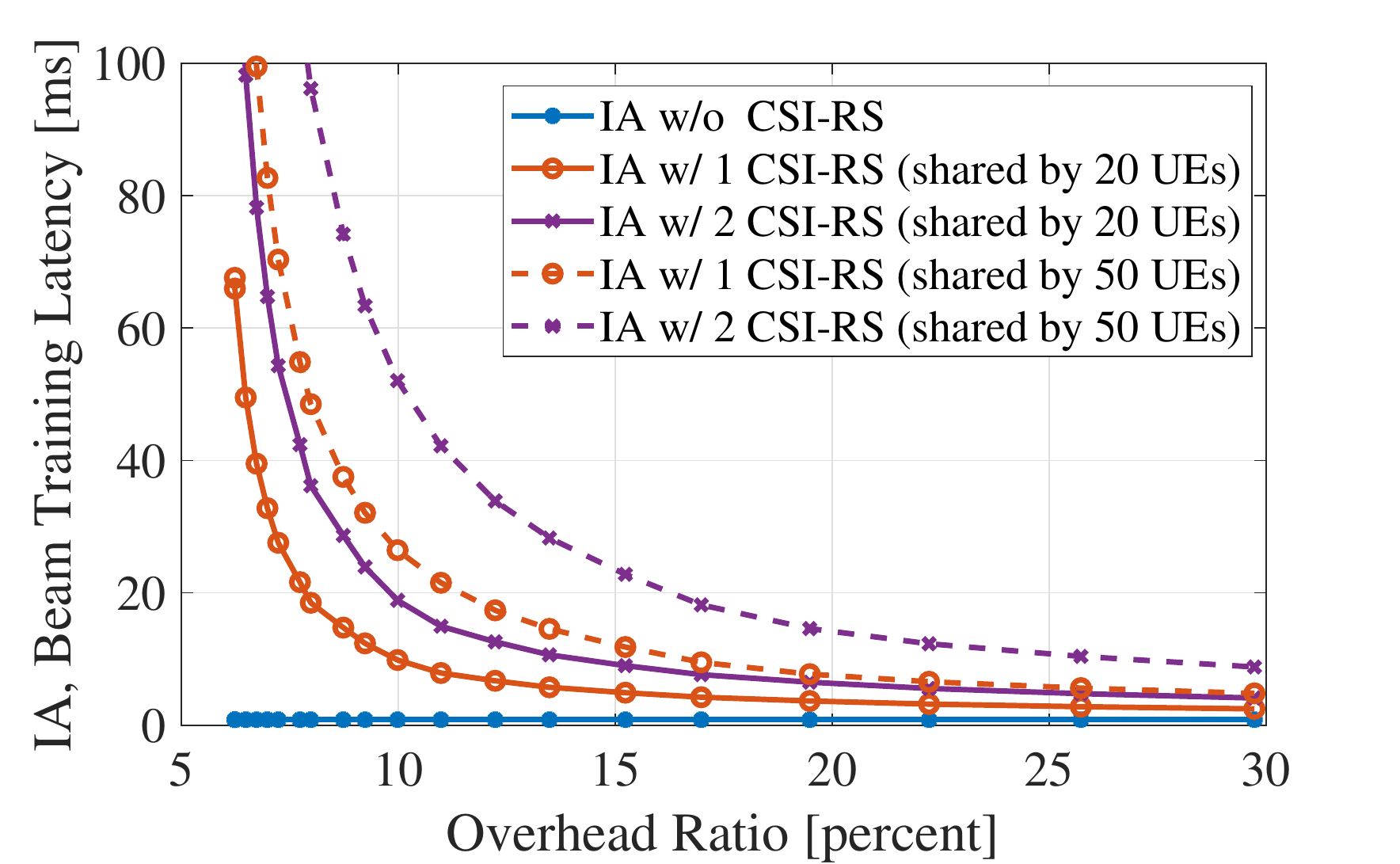}%
}
\end{tabular}
\vspace{0mm}
\caption{Initial access and beam training of proposed and benchmark approaches evaluated in 3D outdoor UMi network using 28GHz mmMAGIC channel model \cite{mmMAGIC_model}. The trade off between post-training SNR in the data phase, required overhead, and access latency are also studied.}
\label{fig:overall_evaluation}
\end{figure*}

Fig.~\ref{fig:overall_evaluation} (a) illustrates the network setting implemented in QuaDRiGa. We simulate the performance of typical UEs distributed in two planes, with different distance towards the pico cell mmW BS. We present the following findings based on Fig.~\ref{fig:overall_evaluation} (b), which shows the cumulative distribution function (CDF) of post-training beam steering SNR. 
Firstly, the proposed approach provides comparable performance to DIA with $N_{\text{train}} = 2$ CSI-RS. In fact, in LOS, both approaches closely achieve beam steering towards true LOS path. Although the SNR seems excessively high in LOS, this implies that the transmit power can be reduced to save power.
Secondly, DIA with less than $N_{\text{train}} = 2$ CSI-RS has compromised SNR performance. This drawback is intuitive because wide sounding sector beam fails to extract precise angle information. The SNR improvement of using higher $N_{\text{train}}$ is more significant in LOS.
Thirdly, although the proposed approach is tailored for sparse channels and presence of phase measurement error due to CFO, it is robust in NLOS scenarios where channel sparsity is compromised and practical phase noise occurs. Admittedly, the algorithm has a certain chance to completely fail when NLOS UEs are distributed in the second plane. However, in these cases the counterparts based on DIA and CSI-RS training cannot do much better job either. In fact, they have lower probability to reach post-training beam steering SNR above $0$ dB compared to the proposed approach.

The overhead and initial access latency savings of the proposed approach are significant, since it does not require CSI-RS, as shown in Fig.~\ref{fig:overall_evaluation} (c). As explained in Section~\ref{sec:comparison_analysis}, for DIA based approaches when number of UEs in the network increases, the latency increases dramatically due to CSI-RS scheduling. Increasing the density of CSI-RS effectively reduces latency, but it results in increased overhead. The proposed approach relies on advanced signal processing to digitally conduct beam training and avoids requesting CSI-RS after initial access. In summary, up to two order of magnitudes saving in initial access latency is reached as compared to DIA.


\subsection{Baseband processing requirements}
Using the simulation parameters in Table~\ref{tab:simulation_setting} to evaluate required operations in Table~\ref{tab:operations}, the baseband resource of the proposed method are in the same order of magnitude with DIA, i.e., $(P\Nb+PG_{\text{d}}+3MG_{\tx}G_{\rx})/(P\Nb) \approx 7.2$. There are two reasons for this finding. Firstly, exhaustive PSS correlation filter (\ref{eq:ZC_correlation}) is extreme  computational demanding in IA. This filter\footnote{In fact, $N_{\text{PSS}}$ filters are required for cell ID identification purpose.} is required by IA regardless of sounding beam design. Secondly, the proposed approach sequentially estimates parameters and avoids multi-dimensional grid search.



%
%
\section{Discussion on Open issues}
\label{sec:discussion}
In this section, we discuss relevant issues in practical implementation of compressive IA and beam training. 

\textit{Required a priori knowledge:} Firstly, this work assumes coarse timing is available. It would be also important to study the case when timing is completely unknown, i.e., there is no a priori information about the range of $\STO$ in (\ref{eq:rx_signal_raw}), which could cause SS burst index misalignment to occur. Secondly, the compressive approach requires precise information about the sounding beam pattern $\tilde{\mathbf{a}}_k$ in (\ref{eq:AoA_AoD_est_start}). As a results, array geometry and sounding codebooks of both BS and UE need to be known a priori. This raises new challenges in communication protocol design to effectively incorporate this information. It also requires an increase in baseband operations if all dictionaries need to be computed on-the-fly. Further, mmW testbed experiments in \cite{SAHA201977} showed that the measured beam patterns commonly have mismatches with patterns predicted by codebook and array geometry model. Future research should address these impairments.

\textit{Channel sparsity:} The efficacy of compressive approach is affected by the sparsity level in AoAs, AoDs, and multipath delays. Sparsity is endorsed by various mmW channel measurement campaigns, and urban NLOS, which is known with infavorable sparsity, is tested in this work. However, severely rich scattering situation are modeled from standard perspective, e,g, there are up to $L=20$ multipath clusters in the 3GPP specified mmW channel \cite{3GPP_model}. It is important for system that utilize CS-based approach to flexibly handle situation when channel sparsity disappears.

\textit{Array architecture:} This work focuses on the scenario where UE uses a single RF-chain to process a single stream of IA signals. This allows other RF-chains, if available at BS or UE, to operate in the band of data communication during IA. Since \cite{7161389} shows that the hybrid analog/digital array and fully digital array are advantageous for DIA, it would be interesting to investigate benefits of compressive IA and beam training algorithm when they are adapted to utilize multiple RF-chain. 

\textit{MIMO Multiplexing:} The proposed beam training is compatible with multiuser multiplexing for hybrid array architecture. In fact, multiplexing designs \cite{7160780,7914742} rely on each RF-chain and corresponding analog beamformer to provide adequate post-BF SNR, and use the digital baseband processing to handle multi-beam interference. However, as mentioned in \textit{Remark 1}, the comparison with channel estimation based approaches, i.e., estimation of the entire wideband channel or its covariance during CSI-RS for optimal MIMO processing, is rarely investigated.

\textit{Phase coherency:} To date, there is no coherent CS-based beam training prototype reported in mmW band. The only notable prototype \cite{7146023} operates at 8GHz with two phased arrays synchronized by cabled reference clock. In addition to CFO, as emphasized in this work, the phase noise can also severely degrade coherency among channel observations. The phase noise detrimental impact becomes more severe with increased carrier frequency. Proper phase noise compensation as well as non-coherent CS-based beam training \cite{Rasekh_noncoherentCS_ACM_2017,UCSB_noncoherentCS_arxiv_1801,Hassanieh:2018:FMW:3230543.3230581} are naturally immune to phase error and are worth investigation.

%
%
\section{Conclusions}
\label{sec:Conclusion}
In this work, quasi-omni pseudorandom sounding beam is proposed for the mmW initial access, synchronization, and beam training. We design associated signal processing algorithm based on the proposed sounding beam structure that is compatible with 5G-NR frame format. We provide theoretical analysis of cell discovery rate and beam training performance, and evaluate them via simulations using the mmW hardware and urban channel models from the literature that are supported by measurements. The results showcase that the proposed approach provides comparable performance to the state-of-the-art directional cell search for initial discovery, but provides significantly more accurate angle estimation during initial beam training. This advantage holds true across different propagation condition (LOS/NLOS) and UE-BS distance at 28 GHz band. Due to the saving of additional radio resource (CSI-RS) for beam refinement, the proposed approach reduces up to two order of magnitude access latency compared to the directional initial access when the same signaling overhead and post-training beam steering SNR are targeted.

All numerical results are reproducible with scripts in \cite{hyan_git}.

\appendix

\subsection{Initial discovery performance}
\label{appendix:detection_derivation_wo_STO}
The noise after correlation $\tilde{z}[n] = \frac{1}{P} \sum_{k=0}^{P-1}(\mathbf{w}^{\hermitian}[n+k]\mathbf{z}[n+k])s_{\text{zc}}^{*}[k]$ is $\mathcal{NC}(0,\sigman^2/P)$. Thus $|\tilde{z}[n]|^2$ is Chi-Square distributed with degree-of-freedom 2, mean $\sigman^2/P$, and variance $\sigman^4/P^2$. 
We denote detection statistic in PT and NT scenario under $\mathcal{H}_0$ and $\mathcal{H}_1$ as denoted as $\gamma_{\text{PT},0}$, $\gamma_{\text{PT},1}$, $\gamma_{\text{NT},0}$, $\gamma_{\text{NT},1}$, respectively, and find their distribution.

$\gamma_{\text{PT},0}$ is the sum of squared $\Nc M$ realizations of $\tilde{z}[n]$ divided by $M$, thus central limit theory (CLT) applies. The distribution of $\gamma_{\text{PT},0}$ is $\mathcal{N}(\mu_{\text{PT},0}, \sigma_{\text{PT},0})$, where   $\mu_{\text{PT},0} =  \Nc\sigman^2/P$ and $\sigma_{\text{PT},0} =  \sqrt{\Nc\sigman^4/(P^2M)}$, respectively. As a result, the optimal detection threshold that reaches target false alarm rate $P^{\star}_{\text{FA}}$ is given by (\ref{eq:TH_w_STO}). 
Similarly, the detection statistic under $\mathcal{H}_0$ with TO is denoted as $\gamma_{\text{NT},0}$. It is the maximum operation with degrees of freedom $\epsilon_{\text{T,max}}$ of $\gamma_{\text{PT},0}$. With large $\epsilon_{\text{T,max}}$, $\gamma_{\text{NT},0}$ follows extreme value distribution, \textit{Gumbel Distribution}, where the mean and standard deviation are 
$\mu_{\text{NT},0} = \mu_{\text{PT},0}+\sigma_{\text{PT},0}\Q^{-1}\left(1/\epsilon_{\text{T,max}}\right) \text{ and }$ and $\sigma_{\text{NT},0} = \sigma_{\text{PT},0}/\Q^{-1}\left(1/\epsilon_{\text{T,max}}\right)$, respectively.
Using its inverse cumulative distribution function, the optimal detection threshold is
$\eta^{\star}_{\text{NT}} = \mu_{\text{NT},0} - (\sqrt{6}\pi)\sigma_{\text{NT},0} \ln\left(-\ln\left(1-P^{\star}_{\text{FA}}\right)\right)$. It gives (\ref{eq:TH_w_STO}) using expressions of $\mu_{\text{NT},0}$, $\sigma_{\text{NT},0}$ and $\sqrt{6}/\pi\approx 0.78$.


Detection statistic $\gamma_{\text{PT},1}$ is the sum of noise energy and signal energy, i.e.,
$\gamma_{\text{PT},1} =\gamma_{\text{PT},0} +  (\sum_{m=1}^{M}\sum_{l=0}^{L}|\tilde{g}_{m,l}\sum_{n=1}^{P}|s_{\text{zc}}[n]|^2e^{j\CFO n}|^2)/(PMN_{\tx}N_{\rx}),$ where $\tilde{g}_{m,l}$ is defined in Section~\ref{sec:signal_rearrangement}. Using the fact $|s_{\text{zc}}[n]|=1$, definition $\kappa(0,\CFO) \triangleq |\sum_{n=1}^{P}e^{j\CFO n}|^2$ in (\ref{eq:kappa_SNR_degradation}), and approximation that different multipaths are resolvable, i.e.,  $p_{\text{c}}(d\Ts-\tau_{l})=1, d\in\mathcal{S}_{\text{d}}$ where $\mathcal{S}_{\text{d}}$ has $L$ integers in range $[0,\Nc-1]$, the above equation becomes $\gamma_{\text{PT},1} = \kappa(0,\CFO) \sum_{m=1}^{M}\zeta_m/M + \gamma_{\text{PT},0}$
where $\zeta_m=\sum_{l=0}^{L}|g_l\mathbf{w}^{\hermitian}_m\mathbf{a}_{\rx}(\phi_l)\mathbf{a}^{\hermitian}_{\tx}(\theta_l)\mathbf{v}_m|^2/(N_{\tx}N_{\rx})$. Using the fact that $\zeta_m$ are mutually independent due to independent $\mathbf{v}_m$ and $\mathbf{w}_m$, the mean and variance of $\zeta_m$ are $\mathbb{E}(\zeta_{m}) = \sum_{l=1}^{L}\left|g_l\right|^2\mathbb{E}\left|\mathbf{w}^{\hermitian}_m\mathbf{a}_{\rx}(\phi_l)\right|^2\mathbb{E}\left|\mathbf{a}^{\hermitian}_{\tx}(\theta_l)\mathbf{v}_m\right|^2/(N_{\tx}N_{\rx}) = \sigma_{\text{g}}^2$
and 
$\var(\zeta_{m}) = (N_{\tx}N_{\rx})^{-2}\sum_{l=1}^{L}\left|g_l\right|^4\mathbb{E}\left|\mathbf{w}^{\hermitian}_m\mathbf{a}_{\rx}(\phi_l)\right|^4\mathbb{E}\left|\mathbf{a}^{\hermitian}_{\tx}(\theta_l)\mathbf{v}_m\right|^4$ $-\sigma_{\text{g}}^4
=\sigma_{\text{g}}^4\left(2-\frac{1}{N_{\tx}}\right)\left(2-\frac{1}{N_{\rx}}\right)-\sigma_{\text{g}}^4 \approx 3\sigma_{\text{g}}^4$, respectively.
The above approximation holds true with typical antenna array sizes $N_{\rx}$ and $N_{\tx}$ in mmW.
Therefore, according to CLT $\gamma_{\text{PT},1}\sim\mathcal{CN}(\kappa(0,\CFO)\sigma_{\text{g}}^2 + \mu_{\text{PT},0},3\kappa^2(0,\CFO)\sigma_{\text{g}}^4/M + \sigma^2_{\text{PT},0})$, which gives the
miss detection probability 
$P_{\text{MD,PT}} = \Q[(\mathbb{E}(\gamma_{\text{PT},1}) - \eta^{\star}_{\text{PT}})/\sqrt{\var(\gamma_{\text{PT},1})}]$, and it equals to (\ref{eq:PMD_w_STO}).
%
%


In NT scenario, 
we make the following approximations: 1) the detection statistic $\gamma_{\text{NT},1}$ corresponds to the correlation peaks for the correct timing $\STO$; 2) the abrupt beamformer changes during $m$-th PSS reception, when present, result in an independent realization of sounding beam $\tilde{\mathbf{w}}_m$. Although the former is not valid with low SNR, the MD rate with typical threshold in such SNR regime already approaches 1. Therefore, impact of such loose approximation is negligible. Based on these assumptions, we evaluate distribution of $\gamma_{\text{NT},1}$ as
$\gamma_{\text{NT},1} = \gamma_{\text{PT},0} + \frac{1}{PMLN_{\tx}N_{\rx}}(\sum_{m=1}^{M}\sum_{l=0}^{L}|\tilde{g}^{(1)}_{m,l}\sum_{n_1=1}^{K-1}|s_{\text{zc}}[n_1]|^2e^{j\CFO n_1}+\tilde{g}^{(2)}_{m,l}\sum_{n_2=K}^{P}|s_{\text{zc}}[n_2]|^2e^{j\CFO n_2}|^2)$
where 
$\tilde{g}^{(1)}_{m,l} =  g_l\mathbf{w}^{\hermitian}_m\mathbf{a}_{\rx}(\phi_l)\mathbf{a}^{\hermitian}_{\tx}(\theta_l)\mathbf{v}_m$ and $\tilde{g}^{(2)}_{m,l} =  g_l\tilde{\mathbf{w}}^{\hermitian}_{m}\mathbf{a}_{\rx}(\phi_l)\mathbf{a}^{\hermitian}_{\tx}(\theta_l)\mathbf{v}_m$
are the post-BF channel gain due to partially overlapped burst window in BS and UE. In other word, $K$ follows (\ref{eq:define_gain_split_K}) and $n_1\in [1,K-1]$ and $n_2\in [K,P]$ are the sample window where $K$ represents the abrupt change in BF. The independent $\mathbf{w}_m$ and $\tilde{\mathbf{w}}_{m}$ lead to uncorrelated $\tilde{g}^{(1)}_{m,l}$ and $\tilde{g}^{(2)}_{m,l}$.
For notational convenience of finding statistic of $\gamma_{\text{NT},1}$, we define $\zeta_{m,l}$ as $
\zeta_{m,l} 
\triangleq  (|\tilde{g}^{(1)}_{m,l}\frac{1-e^{jK\CFO}}{1-e^{j\CFO }}+ \tilde{g}^{(2)}_{m,l}\frac{1-e^{j(P-K)\CFO}}{1-e^{j\CFO }}|^2)/(N_{\tx}N_{\rx})$
in $\gamma_{\text{NT},1}$ after simplification with the fact $|s_{\text{zc}}[n]|^2 = 1, \forall n\in \mathcal{S}$ as well as $\sum_{n=1}^{K}e^{j\STO n} = (1-e^{jK\CFO})/(1-e^{j\CFO})$. The mean and variance of $\zeta_{m,l}$ are $\mathbb{E}\left(\zeta_{m,l}\right)= \kappa(\CFO,\STO)\sigma_{\text{g}}^2$, and $\var\left(\zeta_{m,l}\right) \approx  3\sigma^4_{\text{g}}\zeta^2(\CFO,\STO)$
after plugging in definition of $\kappa(\CFO,\STO)$ from (\ref{eq:kappa_SNR_degradation}).
Using CLT and statistic of $\zeta_{m,l}$, $\gamma_{\text{NT},1}\sim\mathcal{CN} (\mu_{\text{PT},0}+ \kappa(\CFO,\STO) \sigma^2_{\text{g}}, \sigma^2_{\text{PT},0}+ 3\sigma^4_{\text{g}}\kappa^2(\CFO,\STO)/M$. The MD rate $P_{\text{MD,NT}} = \Q[(\mathbb{E}(\gamma_{\text{NT},1}) - \eta^{\star}_{\text{NT}})/\sqrt{\var(\gamma_{\text{NT},1})}]$ reduces to (\ref{eq:PMD_w_STO}). 
\subsection{CRLB of joint estimation problem}
\label{app:CRLB}
\begin{table*}
\caption{Elements of fisher information matrix}
\centering
\def\arraystretch{1.75}
\begin{tabular}{|p{0.6cm}|p{7.7cm}|p{0.55cm}|p{7.5cm}|}
\hline 
Symb. & Expressions & Symb. & Expressions\tabularnewline
\hline
$\Phi_{\CFO,\CFO}$ & $\sum_{m=1}^{M}(C_{\text{d2q}}g)\left|\mathbf{w}_m^{\hermitian}{\mathbf{a}}_{\rx}(\phi)\right|^2\left|\mathbf{v}_m^{\hermitian}{\mathbf{a}_{\tx}}(\theta)\right|^2$  & 
$\Phi_{\CFO,\theta}$ & 
$\sum_{m=1}^{M}(C_m g)\left|\mathbf{w}_m^{\hermitian}\mathbf{a}_{\rx}(\phi)\right|^2\Re\left\{\left[\mathbf{v}_m^{\hermitian}\dot{\mathbf{a}_{\tx}}(\theta)\right]\left[\mathbf{v}_m^{\hermitian}{\mathbf{a}_{\tx}}(\theta)\right]\right\}$\tabularnewline
\hline 
$\Phi_{\CFO,\tau}$
& $\Re\left\{\sum_{m=1}^{M}g\left|\mathbf{w}_m^{\hermitian}{\mathbf{a}}_{\rx}(\phi)\right|^2\left|\mathbf{v}_m^{\hermitian}{\mathbf{a}_{\tx}}(\theta)\right|^2\mathbf{f}^{\hermitian}(\tau)\mathbf{F}\dot{\mathbf{Q}}^{\hermitian}_m{\mathbf{Q}}_m\mathbf{F}^{\hermitian}\dot{\mathbf{f}}(\tau)\right\}$
& $\Phi_{\CFO,\alpha} $
& $\sum_{m=1}^{M}\left[C_{\text{dq},m}\Re\left(g\right)\right]\left|\mathbf{w}_m^{\hermitian}{\mathbf{a}}_{\rx}(\phi)\right|^2\left|\mathbf{v}_m^{\hermitian}{\mathbf{a}_{\tx}}(\theta)\right|^2$
\tabularnewline
\hline
$\Phi_{\CFO,\beta}$
& $\sum_{m=1}^{M}\left[C_{\text{dq},m}\Im\left(g\right)\right]\left|\mathbf{w}_m^{\hermitian}{\mathbf{a}}_{\rx}(\phi)\right|^2\left|\mathbf{v}_m^{\hermitian}{\mathbf{a}_{\tx}}(\theta)\right|^2$
& $\Phi_{\theta,\theta}$
& $\sum_{m=1}^{M}(P|g|^2)\left|\mathbf{w}_m^{\hermitian}\mathbf{a}_{\rx}(\phi)\right|^2\left|\mathbf{v}_m^{\hermitian}\dot{\mathbf{a}_{\tx}}(\theta)\right|^2$
\tabularnewline
\hline
$\Phi_{\phi,\phi}$
& $\sum_{m=1}^{M}(P|g|^2)\left|\mathbf{w}_m^{\hermitian}\dot{\mathbf{a}}_{\rx}(\phi)\right|^2\left|\mathbf{v}_m^{\hermitian}\mathbf{a}_{\tx}(\theta)\right|^2$
& $\Phi_{\phi,\theta}$
& $\Re\left\{\sum_{m=1}^{M}P|g|^2[\mathbf{w}_m^{\hermitian}{\mathbf{a}}_{\rx}(\phi)][\mathbf{w}_m^{\hermitian}\dot{\mathbf{a}}_{\rx}(\phi)][\mathbf{v}_m^{\hermitian}\dot{\mathbf{a}}_{\tx}(\theta)][\mathbf{v}_m^{\hermitian}\mathbf{a}_{\tx}(\theta)]\right\}$
\tabularnewline
\hline
$\Phi_{\phi,\tau}$
& $\sum_{m=1}^{M}C_{\text{df},m}|g|^2\Re\left\{[\mathbf{w}_m^{\hermitian}\dot{\mathbf{a}}_{\rx}(\phi)][\mathbf{w}_m^{\hermitian}{\mathbf{a}}_{\rx}(\phi)]\right\}\left|\mathbf{v}_m^{\hermitian}\mathbf{a}_{\tx}(\theta)\right|^2$
& $\Phi_{\phi,\alpha}$
& $\Re\left\{\sum_{m=1}^{M}Pg[\mathbf{w}_m^{\hermitian}\dot{\mathbf{a}}_{\rx}(\phi)][\mathbf{w}_m^{\hermitian}{\mathbf{a}}_{\rx}(\phi)]|\mathbf{v}_m^{\hermitian}\mathbf{a}_{\tx}(\theta)|^2\right\}$
\tabularnewline
\hline
$\Phi_{\theta,\alpha}$
& $\Re\left\{\sum_{m=1}^{M}Pg|\mathbf{w}_m^{\hermitian}{\mathbf{a}}_{\rx}(\phi)|^2[\mathbf{v}_m^{\hermitian}\dot{\mathbf{a}}_{\tx}(\theta)][\mathbf{v}_m^{\hermitian}\mathbf{a}_{\tx}(\theta)]\right\}$
& $\Phi_{\phi,\beta}$
& $\Re\left\{\sum_{m=1}^{M}jgP[\mathbf{w}_m^{\hermitian}\dot{\mathbf{a}}_{\rx}(\phi)][\mathbf{w}_m^{\hermitian}{\mathbf{a}}_{\rx}(\phi)]|\mathbf{v}_m^{\hermitian}\mathbf{a}_{\tx}(\theta)|^2\right\}$
\tabularnewline
\hline
$\Phi_{\theta,\beta}$
& $\Re\left\{\sum_{m=1}^{M}jPg|\mathbf{w}_m^{\hermitian}{\mathbf{a}}_{\rx}(\phi)|^2[\mathbf{v}_m^{\hermitian}\dot{\mathbf{a}}_{\tx}(\theta)][\mathbf{v}_m^{\hermitian}\mathbf{a}_{\tx}(\theta)]\right\}$
& $\Phi_{\tau,\tau}$
& $\Re\left\{\sum_{m=1}^{M}|g|^2|\mathbf{w}_m^{\hermitian}{\mathbf{a}}_{\rx}(\phi)|^2|\mathbf{v}_m^{\hermitian}\mathbf{a}_{\tx}(\theta)|^2\left[\dot{\mathbf{f}}^{\hermitian}(\tau)\mathbf{Q}_m^{\hermitian}\mathbf{Q}_m\dot{\mathbf{f}}(\tau)\right]\right\}$
\tabularnewline
\hline
$\Phi_{\tau,\alpha}$
& $\Re\left\{\sum_{m=1}^{M}g|\mathbf{w}_m^{\hermitian}{\mathbf{a}}_{\rx}(\phi)|^2|\mathbf{v}_m^{\hermitian}\mathbf{a}_{\tx}(\theta)|^2\left[\dot{\mathbf{f}}^{\hermitian}(\tau)\mathbf{Q}_m^{\hermitian}\mathbf{Q}_m\mathbf{f}(\tau)\right]\right\}$
& $\Phi_{\tau,\beta}$
& $\Re\left\{\sum_{m=1}^{M}jg|\mathbf{w}_m^{\hermitian}{\mathbf{a}}_{\rx}(\phi)|^2|\mathbf{v}_m^{\hermitian}\mathbf{a}_{\tx}(\theta)|^2\left[\dot{\mathbf{f}}^{\hermitian}(\tau)\mathbf{Q}_m^{\hermitian}\mathbf{Q}_m\mathbf{f}(\tau)\right]\right\}$
\tabularnewline
\hline
$\Phi_{\alpha,\alpha}$
& $\sum_{m=1}^{M}P|\mathbf{w}_m^{\hermitian}\mathbf{a}_{\rx}(\phi)|^2|\mathbf{v}_m^{\hermitian}\mathbf{a}_{\tx}(\theta)|^2$
& $\Phi_{\beta,\beta}$
& $-\sum_{m=1}^{M}P|\mathbf{w}_m^{\hermitian}\mathbf{a}_{\rx}(\phi)|^2|\mathbf{v}_m^{\hermitian}\mathbf{a}_{\tx}(\theta)|^2$
\tabularnewline
\hline
\end{tabular}
\label{tab:CRLB}
\end{table*}

The FIM has the following form
\begin{align*}
\mathbf{J} = \frac{1}{\sigman^2}
\begin{bmatrix}
\Phi_{\CFO,\CFO} & \Phi_{\CFO,\theta} & \Phi_{\CFO,\phi} & \Phi_{\CFO,\tau} & \Phi_{\CFO,\alpha} & \Phi_{\CFO,\beta}\\
\Phi_{\theta,\CFO}  & \Phi_{\theta,\theta} & \Phi_{\theta,\phi} &  \Phi_{\theta,\tau} & \Phi_{\theta,\alpha} & \Phi_{\theta,\beta}\\
\Phi_{\phi,\CFO}  & \Phi_{\phi,\theta} & \Phi_{\phi,\phi} &  \Phi_{\phi,\tau} & \Phi_{\phi,\alpha}& \Phi_{\phi,\beta}\\
\Phi_{\tau,\CFO} & \Phi_{\tau,\theta} & \Phi_{\tau,\phi} &  \Phi_{\tau,\tau} & \Phi_{\tau,\alpha} & \Phi_{\tau,\beta}\\
\Phi_{\alpha,\CFO} & \Phi_{\alpha,\theta} & \Phi_{\alpha,\phi} &  \Phi_{\alpha,\tau} & \Phi_{\alpha,\alpha} & 0 \\
\Phi_{\beta,\CFO} & \Phi_{\beta,\theta} & \Phi_{\beta,\phi} &  \Phi_{\beta,\tau} &  0 &\Phi_{\beta,\beta} \\
\end{bmatrix}
\label{eq:FIM}
\end{align*}
where $\Phi_{x,x}$ denotes for $\Phi_{x,x} = \partial^2 L(\mathbf{y};\boldsymbol{\xi})/\partial x \partial y=(\partial L(\mathbf{x}(\boldsymbol{\xi})/\partial x)^{\hermitian}(\partial L(\mathbf{x}(\boldsymbol{\xi}))/\partial y)$. 
The exact expressions of each elements in FIM are summarized in Table~\ref{tab:CRLB}, where for notational convenience the following matrices are defined. The derivative over CFO matrix is a diagonal matrix whose $p$-th diagonal element is $[\dot{\mathbf{Q}}_m]_{p,p} = j[(m-1)\Nb +(p-1)]e^{j\CFO[(m-1)\Nb +(p-1)]}$. The vector  $\dot{\mathbf{f}} = \partial \mathbf{f}(\tau)/\partial \tau$ whose $p$-th element is $[\dot{\mathbf{f}}]_p = 
j2\pi(p-1)\Ts e^{j2\pi(p-1)\CFO \Ts}$
Other expression in Table~\ref{tab:CRLB} include $\mathbf{f}^{\hermitian}(\tau)\mathbf{F}^{\hermitian}\mathbf{Q}_m^{\hermitian}\mathbf{Q}_m\mathbf{F}\mathbf{f}(\tau) = P, \forall m$, $C_{\text{df}} 
= \sum_{p=0}^{P-1} 2\pi p \Ts = (P-2)(P-1)\pi \Ts$ , $C_{\text{dq},m} \triangleq {\mathbf{f}}^{\hermitian}(\tau)\mathbf{F}^{\hermitian}\dot{\mathbf{Q}}_m^{\hermitian}\mathbf{Q}_m\mathbf{F}\mathbf{f}(\tau)  
=(m-1)\Tb + \frac{(P-2)(P-1)\Ts}{2},$ 
, and $C_{\text{d2q},m} = \sum_{p=0}^{P-1}\left[(m-1)\Tb+p\Ts\right]^2$.

\bibliographystyle{IEEEtran}
\bibliography{IEEEabrv,references}

\end{document}